\documentclass[prl,twocolumn,preprintnumbers,amsmath,amssymb,superscriptaddress]{revtex4-1}
\usepackage{graphicx, epsf,bm,amsmath}% Include figure files
\usepackage{subfigure,xcolor}
\usepackage[resetlabels]{multibib}
\newcommand{\br}{{\bf r}}

\begin{document}
\title{Weyl semimetal to metal phase transitions driven by quasiperiodic potentials}
\author{J. H. Pixley}
\affiliation{Department of Physics and Astronomy, Center for Materials Theory, Rutgers University, Piscataway, NJ 08854 USA}
\affiliation{Condensed Matter Theory Center and the Joint Quantum Institute, Department of Physics, University of Maryland, College Park, Maryland 20742-4111 USA}
\author{Justin H. Wilson}
\affiliation{Institute of Quantum Information and Matter and Department of Physics, California Institute of Technology, Pasadena, California 91125 USA}
\author{David A. Huse}
\affiliation{Physics Department, Princeton University, Princeton, NJ 08544 USA}
\author{Sarang Gopalakrishnan}
\affiliation{Department of Engineering Science and Physics, CUNY College of Staten Island, Staten Island, NY 10314 USA and Initiative for the Theoretical Sciences, CUNY Graduate Center, New York, NY 10016 USA}
\date{\today}

\begin{abstract}
We explore the stability of three-dimensional Weyl and Dirac semimetals subject to quasiperiodic potentials. We present numerical evidence that the semimetal is stable for weak quasiperiodic potentials, despite being unstable for weak random potentials. As the quasiperiodic potential strength increases, the semimetal transitions to a metal, then to an ``inverted'' semimetal, and then finally to a metal again. The semimetal and metal are distinguished by the density of states at the Weyl point, as well as by level statistics, transport, and the momentum-space structure of eigenstates near the Weyl point. 
The critical properties of the transitions in quasiperiodic systems differ from those in random systems: we do not find a clear critical scaling regime in energy; instead, at the quasiperiodic transitions, the density of states appears to jump abruptly (and discontinuously to within our resolution).
\end{abstract}

%% Pacs #s
%71.10.Hf	 Non-Fermi-liquid ground states, electron phase diagrams and phase transitions in model systems
%72.80.Ey	III-V and II-VI semiconductors
%73.43.Nq Quantum phase transitions (see also 64.70.Tg Quantum phase transitions in equations of state, phase equilibria and phase transitions)
%72.15.Rn	Localization effects (Anderson or weak localization)
%\pacs{71.10.Hf,72.80.Ey,73.43.Nq,72.15.Rn}

\maketitle

Disorder qualitatively modifies the properties of materials in contexts ranging from spin glasses~\cite{Fischer-1991} to the quantum Hall effect~\cite{Prange-1987}. A striking consequence of disorder in quantum systems is the localization of excitations~\cite{Anderson-1958} and the resulting lack of transport~\cite{Lee-1985, Nandkishore-2015}. While disorder causes localization, it is not a \emph{necessary} condition for localization: deterministic quasiperiodic potentials (QPs) can also support localized excitations~\cite{Azbel-1979,Aubry-1980} but differ from uncorrelated disorder in at least two crucial respects. First, QPs have stable delocalized states even in one-dimension~\cite{Azbel-1979,Aubry-1980} and (unlike disordered systems in any dimension) can exhibit ballistic transport~\cite{sokoloff}. Second, QPs lack large-scale fluctuations, so the rare-region ``Griffiths'' effects that sometimes dominate the behavior of disordered systems~\cite{Griffiths-1969,McCoy-1969} are absent. 
These distinctions are of practical relevance, since experiments with ultracold atoms often use quasiperiodic potentials as an easy-to-implement proxy for randomness~\cite{Roati-2008}. 

The present work addresses a system in which the distinction between quasiperiodicity and randomness is central to the physics, specifically, Weyl semimetals~\cite{Armitage-2017} subject to QPs. In the \emph{random} case, transport at energies near $E = 0$ (i.e., the Weyl point) is anomalous because of the interplay between disorder and the vanishing density of states (DOS)
\cite{Fradkin-1986,Goswami-2011,Kobayashi-2014,Brouwer-2014,Bitan-2014,*Bitan-2016,Nandkishore-2014,Pixley-2015,Altland-2015,Sergey-2015,Leo-2015,Sbierski-2015,Pixley2015disorder,Garttner-2015,Liu-2015,Bera-2015,Shapourian-2015,Altland2-2015,Sergey2-2015,Louvet-2016,Pixley-2016,Pixley2,Sbierski-2017,Pixley-2017,Guararie-2017,Wilson-2017,Wilson-2018}. Disorder is perturbatively irrelevant at the Weyl points~\cite{Fradkin-1986}, suggesting that the ballistic semimetal should be stable to weak disorder (see~\cite{Syzranov-2016} for a recent review from this perspective). However, rare-region effects fill in the zero-energy DOS and destabilize the semimetal for infinitesimal disorder~\cite{Nandkishore-2014,Pixley-2016,Wilson-2018}, although the (so-called) avoided quantum critical point separating the semimetal from the diffusive metal persists as a crossover~\cite{Pixley-2016,Pixley2,Guararie-2017,Wilson-2017}. The random potential plays two roles at this transition: it is both the control parameter for the avoided phase transition and the source of rare regions that destabilize it. To disaggregate these effects, we consider QPs, which lack rare regions. We note that an analogous situation occurs for many-body localization: disorder both drives localization and (through rare-region effects) destabilizes it~\cite{DeRoeck-2017}. The present system potentially offers a more tractable setting with similar phenomena.

\begin{figure}[tb]
\includegraphics[width=\columnwidth]{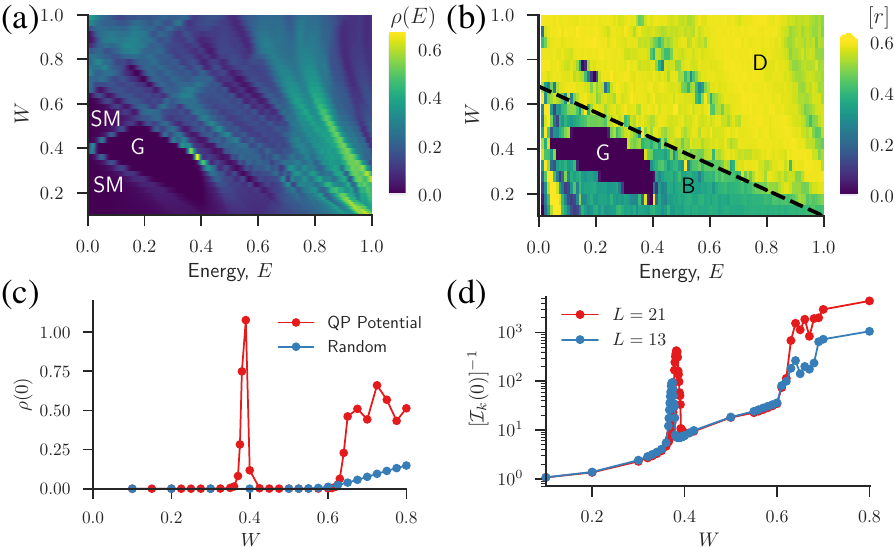}
\caption{Phase diagram in energy ($E$) and QP potential strength ($W$) averaged over 1000 random twists and phases. (a) Density of states (DOS) in the $|E|-W$ plane computed using KPM for $L=55$ and $N_C=2^{10}$. Color represents DOS; we have labeled semi-metallic (SM) and gapped (G) regimes of the model. (b) Level statistics in the $|E|-W$ plane computed from exact diagonalization for $L=13$. The color denotes the average adjacent gap ratio Eq.~\eqref{r}. We have labeled diffusive (D) and ballistic (B) phases. (c) Average zero energy DOS $\rho(0)$ versus $W$ for $L=144$ and $N_C=2^{10}$, from KPM.  (d) Momentum-space IPR for state closest to zero energy, computed for $L = 13, 21$ using Lanczos.} 
\label{fig:1}
\end{figure}

For QPs, we find two types of phases: (i) a semimetal at weak QP strength with ballistic wavefunctions and a vanishing $E = 0$ DOS, and (ii) a diffusive metal at stronger QPs. (We also find an Anderson localized phase for much stronger QPs, but will not focus on the localized phase here.) We present evidence that the semimetal-to-metal transition in this case is sharp and not avoided; its critical properties differ from those of the avoided critical point in the random case. We find a rich phase diagram, featuring a ``mini-band inversion'' transition within the semimetal phase, at which the negative and positive energy states near the Weyl points cross in energy; this crossing is associated with an additional pair of ballistic-to-diffusive transitions around $E = 0$.

\emph{Model, methods, observables}.---
We focus on a three-dimensional model on a simple cubic lattice that represents an inversion-symmetry broken Weyl semimetal
\begin{equation}
H = \sum_{\br, \mu=x,y,z}\frac{1}{2}(it_{\mu}\psi^{\dag}_{\br}\sigma_{\mu}\psi_{\br + \hat{\mu}} + \mathrm{h.c.}) + \sum_{\br} \psi^{\dag}_{\br}V(\br)\psi_{\br}.
\label{eqn:ham}
\end{equation}
$\psi_{\br}$ is a two component spinor, $\sigma_{\mu}$ are the Pauli operators, and the onsite quasiperiodic  potential (QP) is $V(\br)$. We take a three dimensional QP (diagonal in spinor space) $
V(\br) = \sum_{\mu=x,y,z} W_\mu \cos(Q_L r_{\mu}+\phi_{\mu})
$ where each $\phi_{\mu}$ is a random phase sampled between $[0,2\pi]$ that is the same at every site (for the case of a one-dimensional QP see Ref.~\cite{wc2017}). We will also consider the randomized version of the QP potential in which the $\phi_{\mu}$ are random at each lattice site~\cite{Khemani-2017}. This allows us to compare results between these two models at the same $W$ since each site has the same distribution of potentials, and the distinction is whether or not the phases are constant across the system. We consider twisted boundary conditions $t_{\mu}=|t_\mu| \exp(i\theta_{\mu}/L)$, where $\theta_{\mu}$ is randomly sampled between $[0,2\pi]$. We take the linear system size to be given by a Fibbonaci number $L=F_n$ with a wave vector $Q_L=2\pi F_{n-2}/L$ so that as $n\rightarrow \infty$, $Q_L/2\pi\rightarrow 4/(\sqrt{5}+1)^2$. We average over random twists and phases; in the results presented here, we average over 200-1000 samples. 

In this work we consider two slightly distinct models. To locate the critical points in the DOS, we make the simplest choice and set $|t_\mu| = 1, |W_\mu| = W$. 
However, to prevent the threefold symmetry of this model from contaminating level statistics, our studies of level statistics are done on an \emph{anisotropic} model, with broken symmetry in the hopping or both broken symmetry in the hopping and potential; here we take $|t_x| = 1, |t_y| = 0.9, |t_z| = 1.1, |W_x| = W, |W_y| = 0.95 W, |W_z| = 1.1 W$. The models show similar critical behavior, though the (nonuniversal) critical $W$ differs slightly.

We use a combination of numerically exact techniques to study the Hamiltonian in Eq.~\eqref{eqn:ham}. 
To compute the DOS for large systems we use the kernel polynomial method~\cite{Weisse-2006} (KPM). The DOS is 
\begin{equation}
\rho(E) = \left[ L^{-3} \sum\nolimits_i \delta(E - E_i)\right],
\end{equation}
where $E_i$ is the $i$th eigenstate, $[ \dots ]$ denotes a sample average, and $L$ is the linear system size. 
The KPM expands the DOS in Chebyshev polynomials up to an order $N_C$, which is a proxy for energy resolution. We expand the DOS as a Taylor series at low $E$,
$
\rho(E) = \rho(0)+(1/2)\rho''(0)E^2+\cdots
$
and directly compute $\rho''(0)$ with the KPM~\cite{Pixley2}; we expect $\rho''(0)$ to be singular at the semimetal-to-metal transition. 

To study the level statistics and wave functions we use exact diagonalization. For level statistics we compute the adjacent gap ratio 
\begin{equation}\label{r}
r_i \equiv \min(\delta_i,\delta_{i+1})/\max(\delta_i,\delta_{i+1}),
\end{equation}
where $\delta_i = E_{i}-E_{i-1}$ and the eigenvalues have been sorted in ascending order $E_1 < E_2 < \dots < E_N$. Another important metric for us is the momentum-space inverse participation ratio (IPR), defined as:
\begin{equation}\label{iprk}
\mathcal{I}_k(E) \equiv  \left( \sum\nolimits_{\mathbf{k}} |\psi_E(\mathbf{k})|^2 \right)^{-2}   \sum\nolimits_{\mathbf{k}} |\psi_E(\mathbf{k})|^4 
\end{equation}
This quantity probes how much the eigenfunction $\psi_E$ at energy $E$ resembles a plane wave. The ballistic phase is is localized in momentum space and thus $\mathcal{I}_k$ is $L$ independent, whereas in the diffusive phase the wavefunction is spread out in ${\bf k}$ and $\mathcal{I}_k \rightarrow 0$ with increasing $L$. 

We have also computed transport properties~\cite{OSM}, but our results on transport at the transition are inconclusive.  
We use KPM to compute the dynamics of an initially localized wavepacket~\cite{Fehske-2008} in a large system; however, a localized initial state has very little weight near the Weyl points so it is largely insensitive to the transitions of interest here. To focus on the behavior near the Weyl points, we have also computed energy-resolved spectral functions of the local density-density correlation function using exact diagonalization; however, the system sizes accessible here are sufficiently small ($L = 13$) that our results are presumably severely contaminated by finite size effects.

\emph{Phase diagram}.---
We begin by discussing the phase diagram of the model in Eq.~\eqref{eqn:ham} as a function of energy ($E$) and QP strength ($W$), as shown in Fig.~\ref{fig:1}. Unlike disorder, the QP gives rise to an intricate energy-level structure, with mini-bands and hard gaps forming even at relatively weak disorder [Fig.~\ref{fig:1}(a)]. The main features are evident in the color plots of the DOS [Fig.~\ref{fig:1}(a)] and level statistics [Fig.~\ref{fig:1}(b)] as a function of $E$ and $W$.

For small $W$ and $E \approx 0$, the quasiperiodic system behaves like the clean system: the DOS vanishes quadratically at $E = 0$ and all states remain ballistic (i.e., localized in momentum-space). As $W$ is increased, states far away from $E = 0$ become delocalized in momentum-space and develop random-matrix level statistics; we call these energy regimes ``diffusive'' (by analogy with the disordered system). 
At $W \approx 0.15$, minibands around $E = 0$ separate themselves from higher-energy states, and a hard gap appears between the miniband and the higher energy band; we return to this effect below. As $W$ is increased, the positive- and negative-energy minibands merge at $W_m/t = 0.380 \pm 0.001$ at $E=0$ (giving rise to an intermediate, apparently diffusive, metallic phase for $W_m < W \lesssim 0.395 t$ ) and then \emph{cross}: the positive and negative energy minibands change places, and an ``inverted'' semimetal forms~\cite{OSM}. As $W$ increases further, the semimetallic region disappears again at  $W_c \approx 0.6345 \pm 0.001$. For $W \agt W_c$, the DOS at $E = 0$ is finite 
and wave packet dynamics are diffusive~\cite{OSM}. 
$W_c$ in this quasiperiodic model is close to the avoided critical point at $W \approx 0.625$ in the equivalent random model (i.e., the model with random phases at each site in the potential). Level statistics and momentum-space IPR approximately track the DOS---high-DOS regions are typically diffusive and low-DOS regions typically ballistic. 

\begin{figure}[tb]
\begin{center}
\includegraphics[width=\columnwidth]{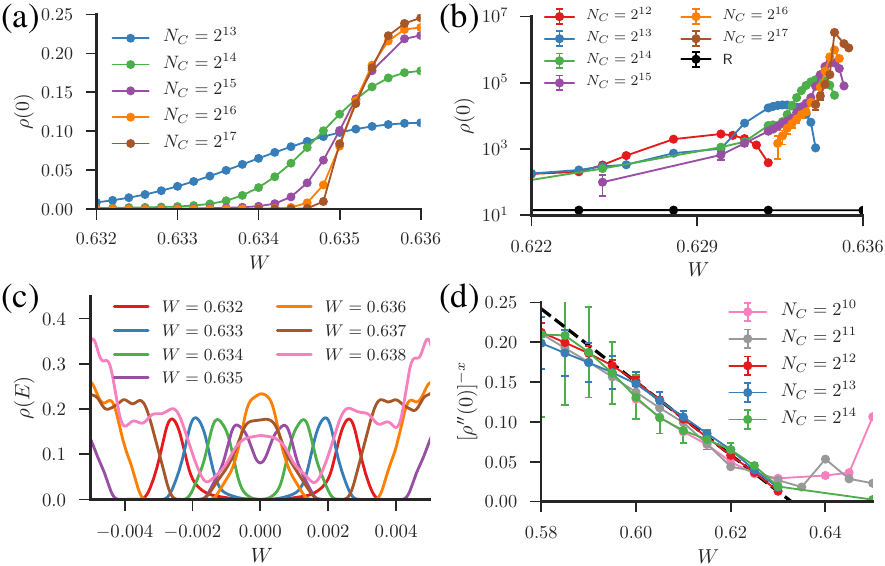}
\caption{Critical behavior of the DOS.
(a) 
$\rho(E = 0)$ vs. $W$ with $L=55$, near the transition at $W_c \approx 0.63$; for various  
KPM expansion orders $N_C$,  
(i.e. different energy resolution). (b)~Second derivative of the DOS, $\rho''(E = 0)$, vs. $W$ with $L=55$, for various $N_C$; $\rho''(0)$ rises steeply with $N_C$, and does not saturate. The black solid line is the data for the equivalent random model (there is a broad peak near $W \approx 0.625$ that looks flat on this scale). (c) 
$\rho(E)$ versus $E$ across the transition, at fixed $L = 55, N_C = 2^{16}$. (d)~
The stability of the scaling regime in $\rho''(E = 0)^{-x}$ versus $W$ with $N_C$ and $L=89$ for $x=0.5$ indicating that $\rho''(E = 0) \sim (W_c - W)^{-2}$, (dashed black line is a linear fit to the data with $N_C=2^{12}$).
}
\label{criticalDOS}
\end{center}
\end{figure}

\emph{Critical properties at $W_c$}.---We now discuss the critical properties of the transition at $W_c$ (Fig.~\ref{criticalDOS}).
 The DOS near $E = 0$, on the semimetallic side ($0.395 t < W< W_c$), goes like $\rho(E)\sim E^2$ and very close to $W_c$ we find $\rho(E) \sim E^2 (W_c - W)^{-\beta}$, with $\beta = 2 \pm 0.8$.
On the ``metallic'' side, the DOS grows rapidly, but we cannot resolve a clear power-law regime. 
In addition, the crossover energy scale at which the low-energy $\rho(E) \sim E^2$ behavior ends appears to shrink linearly with $W_c - W$.
In contrast with the random model, which has a clear critical energy window for which $\rho(E) \sim |E|$, the quasiperiodic model shows no clear scaling other than $\rho(E) \sim E^2$ at the lowest energies in the semimetal. 
The simplest way to account for these observations is if the critical point itself has a nonzero DOS, i.e., at $E=0$ in the infinite system the DOS is discontinuous at the the transition. Our observations are consistent with this scenario; however, we cannot rule out the possibility that the zero-energy DOS instead grows continuously but extremely rapidly.

To identify the non-analyticity of the DOS at $W_c$ we study the dependence of $\rho''(0)$ on the expansion order. For each choice of $N_C \leq 2^{14}$ we converge our data for $\rho(0)$ and $\rho''(0)$ in $L$ so that we know the only rounding is due to $N_C$~\cite{OSM}; however, for $N_C > 2^{15}$
it is infeasible to converge with $L$, so our data are rounded by both $N_C$ and $L$.
The divergence of $\rho''(0)$ in Fig.~\ref{criticalDOS}(b) is striking, reaching $\rho''(0)\sim 10^7$ at $N_C = 2^{17}$, with no sign of saturation. By contrast, in the random problem, the maximum observed $\rho''(0)$ at the avoided transition is $\sim10^3$ (where the peak value was close to saturating)~\cite{Pixley2}. Thus, the transition in the present case appears to be sharp and not avoided. 
This is consistent with both the absence of rare regions and the stability of the ballistic phase, two properties that are common in quasiperiodic systems generally~\cite{Devakul-2017}.

\begin{figure}[tb]
\includegraphics[width=\columnwidth]{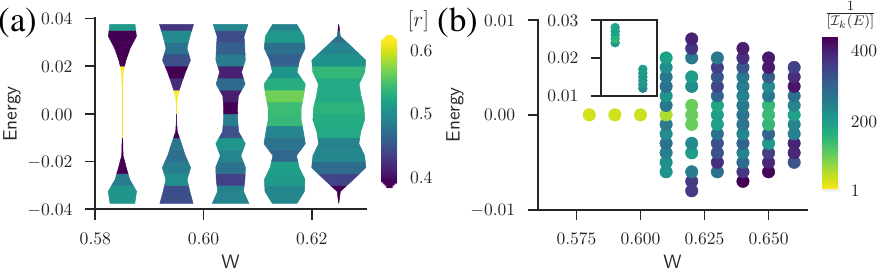}
\caption{Level statistics and momentum-space IPR for $L = 13$ systems near $W_c$; for this system size $W_c \approx 0.61$ averaged over 400 samples. Left: a color plot of 
$[r]$ vs. $E$
and $W$; the color indicates the gap ratio~\eqref{r} in an energy bin, and the size of a dot indicates the DOS in that bin. 
Right: Momentum-space IPR vs. $E$ and $W$; lowest 10 eigenvalues averaged over 100 realizations; inset shows the same quantity at a higher energy range for the $W$ value the inset is above.} 
\label{critls}
\end{figure}

\begin{figure}[tb]
\includegraphics[width=\columnwidth]{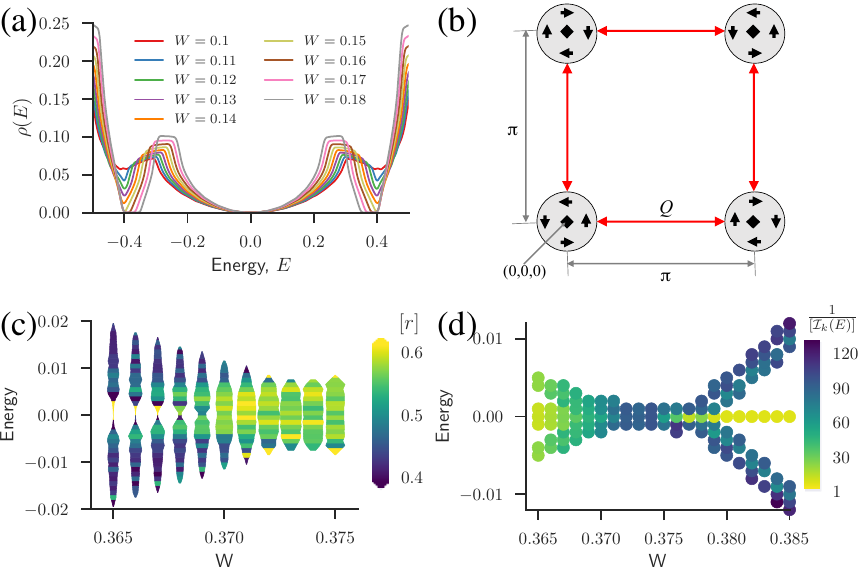}
\caption{Formation of the ``miniband'' and its phase transition. (a)~DOS vs. $E$ at various small $W \approx 0.15$, showing how the miniband detaches from the other states as $W$ increases. (b)~Perturbative structure of the miniband: it forms because the QP hybridizes states from different Weyl cones. Circles denote equal-energy contours in the clean system; thick arrows denote spin textures. (c)~Level statistics [Eq.~\eqref{r}], vs. $E$ and $W$, near the miniband transition, for $L = 13$. (d)~Momentum-space IPR [Eq.~\eqref{iprk}], vs. $E$ and $W$, near the miniband transition, for $L = 13$ (lowest 10 eigenstates pictured).}
\label{miniband}
\end{figure}

\emph{Wavefunctions and level statistics}.---The DOS does not directly tell us whether the system is ballistic or diffusive; a better probe for this is the level statistics parameter $r$ [Eq.~\eqref{r}]. In the ballistic regime, states are localized in momentum space so we expect Poisson level statistics ($[ r ] \approx 0.39$); in the diffusive regime, we expect random-matrix behavior, which (for twisted boundary conditions, which break time-reversal symmetry) should follow the Gaussian unitary ensemble (GUE) ($[ r ] \approx 0.60$). We see the limiting behaviors at small and large $W$; at moderate $W \agt 0.1$ states away from $E = 0$ are mostly diffusive. 
Near $W_c$, the level statistics  
cross over from Poisson-like to GUE-like, though at the accessible system sizes $L = 13$, the level statistics near $E = 0$ is intermediate between Poisson and GUE throughout the 
transition regime. There is considerable inhomogeneity in the level statistics even in the narrow window near zero energy (see Figs.~\ref{critls} and ~\ref{miniband}); notably, at least at the accessible system sizes, states near $E = 0$ appear more random-matrix like %than 
 than higher-DOS regions further from zero energy. The momentum-space IPR largely tracks the level statistics, exhibiting similar heterogeneity; throughout the semimetallic phase, zero-energy states are more tightly localized in momentum space than those away from zero energy. 
Since the momentum-space IPR is straightforward to compute using the Lanczos method, we have been able to look at the spread of IPR for slightly larger systems with $L = 21$; the heterogeneity is more pronounced, but the trend is similar~\cite{OSM}. In general our results suggest that there is a momentum-space delocalization transition that coincides with the DOS transition.

\emph{Miniband transition}.---We now turn to the physics at $W < W_c$ (as shown in Fig.~\ref{miniband}), and discuss some of the fine structure seen inside the semimetal, particularly the miniband ``inversion'' transition. This fine structure is absent in disordered systems. Before exploring how the minibands merge, we first discuss the origin of these minibands; this can be understood using perturbation theory in the QP. The band structure of the clean system consists of eight Weyl points at $(0, 0, 0), (0, 0, \pi), (0, \pi, 0), \ldots (\pi, \pi, \pi)$; for states near the Weyl points, energies are set by the distance in momentum space from the Weyl point. The QP transfers momentum $\sim 0.76 \pi$. Thus, the QP can hybridize a state at $(0.12 \pi, 0, 0)$ with one at $(\pi - 0.12 \pi, 0, 0)$, and so on [see Fig.~\ref{miniband} (b)]. Since these degenerate states belong to Weyl cones with opposite chirality, and are on opposite sides of their Weyl cones, they have the same spin structure and can mix. As the potential becomes stronger, less precisely degenerate states hybridize and this hybridization opens up a gap, separating states that are close to the Weyl nodes (the ``miniband'') from the rest of the band. 

This picture is consistent with our exact diagonalization results on $L=13$. The quasiperiodic approximant for this $L$ has a wavevector $Q_L=2\pi(5/13)$. The estimate above for the characteristic wave-vector of the miniband suggests that only states with momenta that are within %
$(0.5/13) \times 2\pi$ of a Weyl point will contribute to the miniband. There are $54$ such states. 
A miniband consisting of these $54$ states forms at $W \approx 0.15$ [see Fig.~\ref{miniband}(a)]; the miniband is separated from other states by a hard gap.

After the miniband forms, it flattens with increasing $W$, until at $W = W_m \approx 0.38$ the DOS at $E = 0$ fills in (note that for $L = 13$ the apparent $W_m \approx 0.371$ while at $L=21$, $W_m \approx 0.378$). We find $\rho(E)$ becomes non-analytic with a divergence on approach to $W_m$ from both sides, $\rho''(0) \sim |W-W_m|^{-\beta}$ with $\beta = 2 \pm 0.6$~\cite{OSM}. The critical properties of the DOS at this transition are apparently similar to those at $W_c$, but the transition in the level statistics is much clearer in this case [compare Figs.~\ref{critls}(a) and ~\ref{miniband}(c)]. As one approaches the transition, states close to the Weyl point cross over to random-matrix level statistics. Again, the filling-in of the DOS at $E = 0$ coincides (to within our resolution) with the appearance of diffusive states at $E = 0$. 
As one increases $W$ past this point, the minibands separate out and ``invert,'' and their states once again become ballistic. This behavior is again what one would expect perturbatively: the QP mixes states in the miniband with one another only at high orders in perturbation theory, whereas the leading order effect is for each positive (negative) energy state to be pushed down (up) as $W$ increases, leading to an inversion. 
This inversion is driven by $Q_L$ connecting nodes at leading order in perturbation theory; for a smaller $Q_L$ whose leading effect is only intranode hybridization, the inversion disappears~\cite{OSM}.

\emph{Discussion}.---We have provided evidence that Weyl and Dirac semimetals subject to quasiperiodic potentials undergo a true quantum phase transition between a ballistic phase with vanishing DOS at $E = 0$ and a diffusive phase with nonvanishing DOS at $E = 0$. 
We see no indications that the critical point is avoided: the DOS appears strongly nonanalytic, with no sign of intrinsic rounding. The transition affects the DOS, the level statistics, and the structure of wavefunctions at once, to within our resolution: that these should coincide is not \emph{a priori} obvious, as a ballistic phase with finite DOS at $E = 0$ is also possible in principle, and would be natural at sufficiently low DOS. The numerical evidence, however, suggests a discontinuous or at least very steep rise of the DOS at the transition. 
In further contrast to the random case, the DOS lacks a critical energy window.  
In addition to this transition, we found a range of phenomena at weaker quasiperiodic potentials, including the formation of minibands and hard band gaps, and a pair of semimetal-to-metal transitions at which the positive and negative energy minibands merge and go through each other. These miniband transitions are natural in quasiperiodic systems, but their precise location depends on the wavevector $Q_L$.

The most natural experimental setting for exploring the effects studied here is that of ultracold atoms, where Weyl points have already been introduced \cite{jiang_tunable_2012,sun_topological_2012,dubcek_weyl_2015,li_exotic_2015}. The semimetal-to-metal transition can be readily studied in such systems by standard spectroscopic methods (which reveal the DOS) or time-of-flight imaging: the momentum distributions of the ballistic and diffusive phases will 
be quite different. Such systems also provide the possibility of studying interaction effects in Weyl semimetals, and their interplay with quasiperiodic potentials.

\acknowledgements{\emph{Acknowledgements}: We thank Trithep Devakul for useful discussions and Elio K\"onig for comments on a draft. 
The authors are grateful for support from the Laboratory for Physical Sciences (J.~P.), the Air Force Office for Scientific Research (J.~W.), 
and the NSF under DMR-1653271 (S.~G.).
The authors acknowledge the University of Maryland supercomputing resources (http://hpcc.umd.edu), the Beowulf cluster at the Department of Physics and Astronomy of Rutgers University, and the Office of Advanced Research Computing (OARC) at Rutgers, The State University of New Jersey (http://oarc.rutgers.edu) for providing access to the Amarel cluster and associated research computing resources that have contributed to the results reported here. 
}

\bibliography{DSM_RR}

\begin{thebibliography}{51}%
\makeatletter
\providecommand \@ifxundefined [1]{%
 \@ifx{#1\undefined}
}%
\providecommand \@ifnum [1]{%
 \ifnum #1\expandafter \@firstoftwo
 \else \expandafter \@secondoftwo
 \fi
}%
\providecommand \@ifx [1]{%
 \ifx #1\expandafter \@firstoftwo
 \else \expandafter \@secondoftwo
 \fi
}%
\providecommand \natexlab [1]{#1}%
\providecommand \enquote  [1]{``#1''}%
\providecommand \bibnamefont  [1]{#1}%
\providecommand \bibfnamefont [1]{#1}%
\providecommand \citenamefont [1]{#1}%
\providecommand \href@noop [0]{\@secondoftwo}%
\providecommand \href [0]{\begingroup \@sanitize@url \@href}%
\providecommand \@href[1]{\@@startlink{#1}\@@href}%
\providecommand \@@href[1]{\endgroup#1\@@endlink}%
\providecommand \@sanitize@url [0]{\catcode `\\12\catcode `\$12\catcode
  `\&12\catcode `\#12\catcode `\^12\catcode `\_12\catcode `\%12\relax}%
\providecommand \@@startlink[1]{}%
\providecommand \@@endlink[0]{}%
\providecommand \url  [0]{\begingroup\@sanitize@url \@url }%
\providecommand \@url [1]{\endgroup\@href {#1}{\urlprefix }}%
\providecommand \urlprefix  [0]{URL }%
\providecommand \Eprint [0]{\href }%
\providecommand \doibase [0]{http://dx.doi.org/}%
\providecommand \selectlanguage [0]{\@gobble}%
\providecommand \bibinfo  [0]{\@secondoftwo}%
\providecommand \bibfield  [0]{\@secondoftwo}%
\providecommand \translation [1]{[#1]}%
\providecommand \BibitemOpen [0]{}%
\providecommand \bibitemStop [0]{}%
\providecommand \bibitemNoStop [0]{.\EOS\space}%
\providecommand \EOS [0]{\spacefactor3000\relax}%
\providecommand \BibitemShut  [1]{\csname bibitem#1\endcsname}%
\let\auto@bib@innerbib\@empty
%</preamble>
\bibitem [{\citenamefont {Fischer}\ and\ \citenamefont
  {Hertz}(1991)}]{Fischer-1991}%
  \BibitemOpen
  \bibfield  {author} {\bibinfo {author} {\bibfnamefont {K.}~\bibnamefont
  {Fischer}}\ and\ \bibinfo {author} {\bibfnamefont {J.}~\bibnamefont
  {Hertz}},\ }\href@noop {} {\enquote {\bibinfo {title} {Spin-glasses, volume 1
  of cambridge studies in magnetism},}\ } (\bibinfo {year} {1991})\BibitemShut
  {NoStop}%
\bibitem [{\citenamefont {Prange}\ and\ \citenamefont
  {Girvin}(1987)}]{Prange-1987}%
  \BibitemOpen
  \bibfield  {author} {\bibinfo {author} {\bibfnamefont {R.~E.}\ \bibnamefont
  {Prange}}\ and\ \bibinfo {author} {\bibfnamefont {S.~M.}\ \bibnamefont
  {Girvin}},\ }\href@noop {} {\enquote {\bibinfo {title} {The quantum hall
  effect, graduate texts in contemporary physics},}\ } (\bibinfo {year}
  {1987})\BibitemShut {NoStop}%
\bibitem [{\citenamefont {Anderson}(1958)}]{Anderson-1958}%
  \BibitemOpen
  \bibfield  {author} {\bibinfo {author} {\bibfnamefont {P.~W.}\ \bibnamefont
  {Anderson}},\ }\href@noop {} {\bibfield  {journal} {\bibinfo  {journal}
  {Phys. Rev.}\ }\textbf {\bibinfo {volume} {109}},\ \bibinfo {pages} {1492}
  (\bibinfo {year} {1958})}\BibitemShut {NoStop}%
\bibitem [{\citenamefont {Lee}\ and\ \citenamefont
  {Ramakrishnan}(1985)}]{Lee-1985}%
  \BibitemOpen
  \bibfield  {author} {\bibinfo {author} {\bibfnamefont {P.~A.}\ \bibnamefont
  {Lee}}\ and\ \bibinfo {author} {\bibfnamefont {T.~V.}\ \bibnamefont
  {Ramakrishnan}},\ }\href@noop {} {\bibfield  {journal} {\bibinfo  {journal}
  {Rev. Mod. Phys.}\ }\textbf {\bibinfo {volume} {57}},\ \bibinfo {pages} {287}
  (\bibinfo {year} {1985})}\BibitemShut {NoStop}%
\bibitem [{\citenamefont {Nandkishore}\ and\ \citenamefont
  {Huse}(2015)}]{Nandkishore-2015}%
  \BibitemOpen
  \bibfield  {author} {\bibinfo {author} {\bibfnamefont {R.}~\bibnamefont
  {Nandkishore}}\ and\ \bibinfo {author} {\bibfnamefont {D.~A.}\ \bibnamefont
  {Huse}},\ }\href@noop {} {\bibfield  {journal} {\bibinfo  {journal} {Annu.
  Rev. Condens. Matter Phys.}\ }\textbf {\bibinfo {volume} {6}},\ \bibinfo
  {pages} {15} (\bibinfo {year} {2015})}\BibitemShut {NoStop}%
\bibitem [{\citenamefont {Azbel}(1979)}]{Azbel-1979}%
  \BibitemOpen
  \bibfield  {author} {\bibinfo {author} {\bibfnamefont {M.~Y.}\ \bibnamefont
  {Azbel}},\ }\href {\doibase 10.1103/PhysRevLett.43.1954} {\bibfield
  {journal} {\bibinfo  {journal} {Phys. Rev. Lett.}\ }\textbf {\bibinfo
  {volume} {43}},\ \bibinfo {pages} {1954} (\bibinfo {year}
  {1979})}\BibitemShut {NoStop}%
\bibitem [{\citenamefont {Aubry}\ and\ \citenamefont
  {Andr{\'e}}(1980)}]{Aubry-1980}%
  \BibitemOpen
  \bibfield  {author} {\bibinfo {author} {\bibfnamefont {S.}~\bibnamefont
  {Aubry}}\ and\ \bibinfo {author} {\bibfnamefont {G.}~\bibnamefont
  {Andr{\'e}}},\ }\href@noop {} {\bibfield  {journal} {\bibinfo  {journal}
  {Ann. Israel Phys. Soc}\ }\textbf {\bibinfo {volume} {3}},\ \bibinfo {pages}
  {18} (\bibinfo {year} {1980})}\BibitemShut {NoStop}%
\bibitem [{\citenamefont {Sokoloff}(1985)}]{sokoloff}%
  \BibitemOpen
  \bibfield  {author} {\bibinfo {author} {\bibfnamefont {J.}~\bibnamefont
  {Sokoloff}},\ }\href@noop {} {\bibfield  {journal} {\bibinfo  {journal}
  {Physics Reports}\ }\textbf {\bibinfo {volume} {126}},\ \bibinfo {pages}
  {189} (\bibinfo {year} {1985})}\BibitemShut {NoStop}%
\bibitem [{\citenamefont {Griffiths}(1969)}]{Griffiths-1969}%
  \BibitemOpen
  \bibfield  {author} {\bibinfo {author} {\bibfnamefont {R.~B.}\ \bibnamefont
  {Griffiths}},\ }\href {\doibase 10.1103/PhysRevLett.23.17} {\bibfield
  {journal} {\bibinfo  {journal} {Phys. Rev. Lett.}\ }\textbf {\bibinfo
  {volume} {23}},\ \bibinfo {pages} {17} (\bibinfo {year} {1969})}\BibitemShut
  {NoStop}%
\bibitem [{\citenamefont {McCoy}(1969)}]{McCoy-1969}%
  \BibitemOpen
  \bibfield  {author} {\bibinfo {author} {\bibfnamefont {B.~M.}\ \bibnamefont
  {McCoy}},\ }\href {\doibase 10.1103/PhysRevLett.23.383} {\bibfield  {journal}
  {\bibinfo  {journal} {Phys. Rev. Lett.}\ }\textbf {\bibinfo {volume} {23}},\
  \bibinfo {pages} {383} (\bibinfo {year} {1969})}\BibitemShut {NoStop}%
\bibitem [{\citenamefont {Roati}\ \emph {et~al.}(2008)\citenamefont {Roati},
  \citenamefont {D’Errico}, \citenamefont {Fallani}, \citenamefont {Fattori},
  \citenamefont {Fort}, \citenamefont {Zaccanti}, \citenamefont {Modugno},
  \citenamefont {Modugno},\ and\ \citenamefont {Inguscio}}]{Roati-2008}%
  \BibitemOpen
  \bibfield  {author} {\bibinfo {author} {\bibfnamefont {G.}~\bibnamefont
  {Roati}}, \bibinfo {author} {\bibfnamefont {C.}~\bibnamefont {D’Errico}},
  \bibinfo {author} {\bibfnamefont {L.}~\bibnamefont {Fallani}}, \bibinfo
  {author} {\bibfnamefont {M.}~\bibnamefont {Fattori}}, \bibinfo {author}
  {\bibfnamefont {C.}~\bibnamefont {Fort}}, \bibinfo {author} {\bibfnamefont
  {M.}~\bibnamefont {Zaccanti}}, \bibinfo {author} {\bibfnamefont
  {G.}~\bibnamefont {Modugno}}, \bibinfo {author} {\bibfnamefont
  {M.}~\bibnamefont {Modugno}}, \ and\ \bibinfo {author} {\bibfnamefont
  {M.}~\bibnamefont {Inguscio}},\ }\href@noop {} {\bibfield  {journal}
  {\bibinfo  {journal} {Nature}\ }\textbf {\bibinfo {volume} {453}},\ \bibinfo
  {pages} {895} (\bibinfo {year} {2008})}\BibitemShut {NoStop}%
\bibitem [{\citenamefont {Armitage}\ \emph {et~al.}(2017)\citenamefont
  {Armitage}, \citenamefont {Mele},\ and\ \citenamefont
  {Vishwanath}}]{Armitage-2017}%
  \BibitemOpen
  \bibfield  {author} {\bibinfo {author} {\bibfnamefont {N.}~\bibnamefont
  {Armitage}}, \bibinfo {author} {\bibfnamefont {E.}~\bibnamefont {Mele}}, \
  and\ \bibinfo {author} {\bibfnamefont {A.}~\bibnamefont {Vishwanath}},\
  }\href@noop {} {\bibfield  {journal} {\bibinfo  {journal} {arXiv preprint
  arXiv:1705.01111}\ } (\bibinfo {year} {2017})}\BibitemShut {NoStop}%
\bibitem [{\citenamefont {Fradkin}(1986)}]{Fradkin-1986}%
  \BibitemOpen
  \bibfield  {author} {\bibinfo {author} {\bibfnamefont {E.}~\bibnamefont
  {Fradkin}},\ }\href@noop {} {\bibfield  {journal} {\bibinfo  {journal} {Phys.
  Rev. B}\ }\textbf {\bibinfo {volume} {33}},\ \bibinfo {pages} {3263}
  (\bibinfo {year} {1986})}\BibitemShut {NoStop}%
\bibitem [{\citenamefont {Goswami}\ and\ \citenamefont
  {Chakravarty}(2011)}]{Goswami-2011}%
  \BibitemOpen
  \bibfield  {author} {\bibinfo {author} {\bibfnamefont {P.}~\bibnamefont
  {Goswami}}\ and\ \bibinfo {author} {\bibfnamefont {S.}~\bibnamefont
  {Chakravarty}},\ }\href@noop {} {\bibfield  {journal} {\bibinfo  {journal}
  {Phys. Rev. Lett.}\ }\textbf {\bibinfo {volume} {107}},\ \bibinfo {pages}
  {196803} (\bibinfo {year} {2011})}\BibitemShut {NoStop}%
\bibitem [{\citenamefont {Kobayashi}\ \emph {et~al.}(2014)\citenamefont
  {Kobayashi}, \citenamefont {Ohtsuki}, \citenamefont {Imura},\ and\
  \citenamefont {Herbut}}]{Kobayashi-2014}%
  \BibitemOpen
  \bibfield  {author} {\bibinfo {author} {\bibfnamefont {K.}~\bibnamefont
  {Kobayashi}}, \bibinfo {author} {\bibfnamefont {T.}~\bibnamefont {Ohtsuki}},
  \bibinfo {author} {\bibfnamefont {K.-I.}\ \bibnamefont {Imura}}, \ and\
  \bibinfo {author} {\bibfnamefont {I.~F.}\ \bibnamefont {Herbut}},\
  }\href@noop {} {\bibfield  {journal} {\bibinfo  {journal} {Phys. Rev. Lett.}\
  }\textbf {\bibinfo {volume} {112}},\ \bibinfo {pages} {016402} (\bibinfo
  {year} {2014})}\BibitemShut {NoStop}%
\bibitem [{\citenamefont {Sbierski}\ \emph {et~al.}(2014)\citenamefont
  {Sbierski}, \citenamefont {Pohl}, \citenamefont {Bergholtz},\ and\
  \citenamefont {Brouwer}}]{Brouwer-2014}%
  \BibitemOpen
  \bibfield  {author} {\bibinfo {author} {\bibfnamefont {B.}~\bibnamefont
  {Sbierski}}, \bibinfo {author} {\bibfnamefont {G.}~\bibnamefont {Pohl}},
  \bibinfo {author} {\bibfnamefont {E.~J.}\ \bibnamefont {Bergholtz}}, \ and\
  \bibinfo {author} {\bibfnamefont {P.~W.}\ \bibnamefont {Brouwer}},\
  }\href@noop {} {\bibfield  {journal} {\bibinfo  {journal} {Phys. Rev. Lett.}\
  }\textbf {\bibinfo {volume} {113}},\ \bibinfo {pages} {026602} (\bibinfo
  {year} {2014})}\BibitemShut {NoStop}%
\bibitem [{\citenamefont {Roy}\ and\ \citenamefont
  {Das~Sarma}(2014)}]{Bitan-2014}%
  \BibitemOpen
  \bibfield  {author} {\bibinfo {author} {\bibfnamefont {B.}~\bibnamefont
  {Roy}}\ and\ \bibinfo {author} {\bibfnamefont {S.}~\bibnamefont
  {Das~Sarma}},\ }\href@noop {} {\bibfield  {journal} {\bibinfo  {journal}
  {Phys. Rev. B}\ }\textbf {\bibinfo {volume} {90}},\ \bibinfo {pages} {241112}
  (\bibinfo {year} {2014})}\BibitemShut {NoStop}%
\bibitem [{\citenamefont {Roy}\ and\ \citenamefont
  {Das~Sarma}(2016)}]{Bitan-2016}%
  \BibitemOpen
  \bibfield  {author} {\bibinfo {author} {\bibfnamefont {B.}~\bibnamefont
  {Roy}}\ and\ \bibinfo {author} {\bibfnamefont {S.}~\bibnamefont
  {Das~Sarma}},\ }\href@noop {} {\bibfield  {journal} {\bibinfo  {journal}
  {Phys. Rev. B}\ }\textbf {\bibinfo {volume} {93}},\ \bibinfo {pages} {119911}
  (\bibinfo {year} {2016})}\BibitemShut {NoStop}%
\bibitem [{\citenamefont {Nandkishore}\ \emph {et~al.}(2014)\citenamefont
  {Nandkishore}, \citenamefont {Huse},\ and\ \citenamefont
  {Sondhi}}]{Nandkishore-2014}%
  \BibitemOpen
  \bibfield  {author} {\bibinfo {author} {\bibfnamefont {R.}~\bibnamefont
  {Nandkishore}}, \bibinfo {author} {\bibfnamefont {D.~A.}\ \bibnamefont
  {Huse}}, \ and\ \bibinfo {author} {\bibfnamefont {S.~L.}\ \bibnamefont
  {Sondhi}},\ }\href@noop {} {\bibfield  {journal} {\bibinfo  {journal}
  {Physical Review B}\ }\textbf {\bibinfo {volume} {89}},\ \bibinfo {pages}
  {245110} (\bibinfo {year} {2014})}\BibitemShut {NoStop}%
\bibitem [{\citenamefont {Pixley}\ \emph {et~al.}(2015)\citenamefont {Pixley},
  \citenamefont {Goswami},\ and\ \citenamefont {Das~Sarma}}]{Pixley-2015}%
  \BibitemOpen
  \bibfield  {author} {\bibinfo {author} {\bibfnamefont {J.~H.}\ \bibnamefont
  {Pixley}}, \bibinfo {author} {\bibfnamefont {P.}~\bibnamefont {Goswami}}, \
  and\ \bibinfo {author} {\bibfnamefont {S.}~\bibnamefont {Das~Sarma}},\
  }\href@noop {} {\bibfield  {journal} {\bibinfo  {journal} {Phys. Rev. Lett.}\
  }\textbf {\bibinfo {volume} {115}},\ \bibinfo {pages} {076601} (\bibinfo
  {year} {2015})}\BibitemShut {NoStop}%
\bibitem [{\citenamefont {Altland}\ and\ \citenamefont
  {Bagrets}(2015)}]{Altland-2015}%
  \BibitemOpen
  \bibfield  {author} {\bibinfo {author} {\bibfnamefont {A.}~\bibnamefont
  {Altland}}\ and\ \bibinfo {author} {\bibfnamefont {D.}~\bibnamefont
  {Bagrets}},\ }\href@noop {} {\bibfield  {journal} {\bibinfo  {journal} {Phys.
  Rev. Lett.}\ }\textbf {\bibinfo {volume} {114}},\ \bibinfo {pages} {257201}
  (\bibinfo {year} {2015})}\BibitemShut {NoStop}%
\bibitem [{\citenamefont {Syzranov}\ \emph
  {et~al.}(2015{\natexlab{a}})\citenamefont {Syzranov}, \citenamefont
  {Radzihovsky},\ and\ \citenamefont {Gurarie}}]{Sergey-2015}%
  \BibitemOpen
  \bibfield  {author} {\bibinfo {author} {\bibfnamefont {S.~V.}\ \bibnamefont
  {Syzranov}}, \bibinfo {author} {\bibfnamefont {L.}~\bibnamefont
  {Radzihovsky}}, \ and\ \bibinfo {author} {\bibfnamefont {V.}~\bibnamefont
  {Gurarie}},\ }\href@noop {} {\bibfield  {journal} {\bibinfo  {journal} {Phys.
  Rev. Lett.}\ }\textbf {\bibinfo {volume} {114}},\ \bibinfo {pages} {166601}
  (\bibinfo {year} {2015}{\natexlab{a}})}\BibitemShut {NoStop}%
\bibitem [{\citenamefont {Syzranov}\ \emph
  {et~al.}(2015{\natexlab{b}})\citenamefont {Syzranov}, \citenamefont
  {Gurarie},\ and\ \citenamefont {Radzihovsky}}]{Leo-2015}%
  \BibitemOpen
  \bibfield  {author} {\bibinfo {author} {\bibfnamefont {S.~V.}\ \bibnamefont
  {Syzranov}}, \bibinfo {author} {\bibfnamefont {V.}~\bibnamefont {Gurarie}}, \
  and\ \bibinfo {author} {\bibfnamefont {L.}~\bibnamefont {Radzihovsky}},\
  }\href@noop {} {\bibfield  {journal} {\bibinfo  {journal} {Phys. Rev. B}\
  }\textbf {\bibinfo {volume} {91}},\ \bibinfo {pages} {035133} (\bibinfo
  {year} {2015}{\natexlab{b}})}\BibitemShut {NoStop}%
\bibitem [{\citenamefont {Sbierski}\ \emph {et~al.}(2015)\citenamefont
  {Sbierski}, \citenamefont {Bergholtz},\ and\ \citenamefont
  {Brouwer}}]{Sbierski-2015}%
  \BibitemOpen
  \bibfield  {author} {\bibinfo {author} {\bibfnamefont {B.}~\bibnamefont
  {Sbierski}}, \bibinfo {author} {\bibfnamefont {E.~J.}\ \bibnamefont
  {Bergholtz}}, \ and\ \bibinfo {author} {\bibfnamefont {P.~W.}\ \bibnamefont
  {Brouwer}},\ }\href@noop {} {\bibfield  {journal} {\bibinfo  {journal} {Phys.
  Rev. B}\ }\textbf {\bibinfo {volume} {92}},\ \bibinfo {pages} {115145}
  (\bibinfo {year} {2015})}\BibitemShut {NoStop}%
\bibitem [{\citenamefont {Pixley}\ \emph
  {et~al.}(2016{\natexlab{a}})\citenamefont {Pixley}, \citenamefont {Goswami},\
  and\ \citenamefont {Das~Sarma}}]{Pixley2015disorder}%
  \BibitemOpen
  \bibfield  {author} {\bibinfo {author} {\bibfnamefont {J.~H.}\ \bibnamefont
  {Pixley}}, \bibinfo {author} {\bibfnamefont {P.}~\bibnamefont {Goswami}}, \
  and\ \bibinfo {author} {\bibfnamefont {S.}~\bibnamefont {Das~Sarma}},\
  }\href@noop {} {\bibfield  {journal} {\bibinfo  {journal} {Phys. Rev. B}\
  }\textbf {\bibinfo {volume} {93}},\ \bibinfo {pages} {085103} (\bibinfo
  {year} {2016}{\natexlab{a}})}\BibitemShut {NoStop}%
\bibitem [{\citenamefont {G\"arttner}\ \emph {et~al.}(2015)\citenamefont
  {G\"arttner}, \citenamefont {Syzranov}, \citenamefont {Rey}, \citenamefont
  {Gurarie},\ and\ \citenamefont {Radzihovsky}}]{Garttner-2015}%
  \BibitemOpen
  \bibfield  {author} {\bibinfo {author} {\bibfnamefont {M.}~\bibnamefont
  {G\"arttner}}, \bibinfo {author} {\bibfnamefont {S.~V.}\ \bibnamefont
  {Syzranov}}, \bibinfo {author} {\bibfnamefont {A.~M.}\ \bibnamefont {Rey}},
  \bibinfo {author} {\bibfnamefont {V.}~\bibnamefont {Gurarie}}, \ and\
  \bibinfo {author} {\bibfnamefont {L.}~\bibnamefont {Radzihovsky}},\
  }\href@noop {} {\bibfield  {journal} {\bibinfo  {journal} {Phys. Rev. B}\
  }\textbf {\bibinfo {volume} {92}},\ \bibinfo {pages} {041406} (\bibinfo
  {year} {2015})}\BibitemShut {NoStop}%
\bibitem [{\citenamefont {Liu}\ \emph {et~al.}(2016)\citenamefont {Liu},
  \citenamefont {Ohtsuki},\ and\ \citenamefont {Shindou}}]{Liu-2015}%
  \BibitemOpen
  \bibfield  {author} {\bibinfo {author} {\bibfnamefont {S.}~\bibnamefont
  {Liu}}, \bibinfo {author} {\bibfnamefont {T.}~\bibnamefont {Ohtsuki}}, \ and\
  \bibinfo {author} {\bibfnamefont {R.}~\bibnamefont {Shindou}},\ }\href@noop
  {} {\bibfield  {journal} {\bibinfo  {journal} {Phys. Rev. Lett.}\ }\textbf
  {\bibinfo {volume} {116}},\ \bibinfo {pages} {066401} (\bibinfo {year}
  {2016})}\BibitemShut {NoStop}%
\bibitem [{\citenamefont {Bera}\ \emph {et~al.}(2016)\citenamefont {Bera},
  \citenamefont {Sau},\ and\ \citenamefont {Roy}}]{Bera-2015}%
  \BibitemOpen
  \bibfield  {author} {\bibinfo {author} {\bibfnamefont {S.}~\bibnamefont
  {Bera}}, \bibinfo {author} {\bibfnamefont {J.~D.}\ \bibnamefont {Sau}}, \
  and\ \bibinfo {author} {\bibfnamefont {B.}~\bibnamefont {Roy}},\ }\href@noop
  {} {\bibfield  {journal} {\bibinfo  {journal} {Phys. Rev. B}\ }\textbf
  {\bibinfo {volume} {93}},\ \bibinfo {pages} {201302} (\bibinfo {year}
  {2016})}\BibitemShut {NoStop}%
\bibitem [{\citenamefont {Shapourian}\ and\ \citenamefont
  {Hughes}(2016)}]{Shapourian-2015}%
  \BibitemOpen
  \bibfield  {author} {\bibinfo {author} {\bibfnamefont {H.}~\bibnamefont
  {Shapourian}}\ and\ \bibinfo {author} {\bibfnamefont {T.~L.}\ \bibnamefont
  {Hughes}},\ }\href@noop {} {\bibfield  {journal} {\bibinfo  {journal} {Phys.
  Rev. B}\ }\textbf {\bibinfo {volume} {93}},\ \bibinfo {pages} {075108}
  (\bibinfo {year} {2016})}\BibitemShut {NoStop}%
\bibitem [{\citenamefont {Altland}\ and\ \citenamefont
  {Bagrets}(2016)}]{Altland2-2015}%
  \BibitemOpen
  \bibfield  {author} {\bibinfo {author} {\bibfnamefont {A.}~\bibnamefont
  {Altland}}\ and\ \bibinfo {author} {\bibfnamefont {D.}~\bibnamefont
  {Bagrets}},\ }\href@noop {} {\bibfield  {journal} {\bibinfo  {journal} {Phys.
  Rev. B}\ }\textbf {\bibinfo {volume} {93}},\ \bibinfo {pages} {075113}
  (\bibinfo {year} {2016})}\BibitemShut {NoStop}%
\bibitem [{\citenamefont {Syzranov}\ \emph {et~al.}(2016)\citenamefont
  {Syzranov}, \citenamefont {Ostrovsky}, \citenamefont {Gurarie},\ and\
  \citenamefont {Radzihovsky}}]{Sergey2-2015}%
  \BibitemOpen
  \bibfield  {author} {\bibinfo {author} {\bibfnamefont {S.~V.}\ \bibnamefont
  {Syzranov}}, \bibinfo {author} {\bibfnamefont {P.~M.}\ \bibnamefont
  {Ostrovsky}}, \bibinfo {author} {\bibfnamefont {V.}~\bibnamefont {Gurarie}},
  \ and\ \bibinfo {author} {\bibfnamefont {L.}~\bibnamefont {Radzihovsky}},\
  }\href@noop {} {\bibfield  {journal} {\bibinfo  {journal} {Phys. Rev. B}\
  }\textbf {\bibinfo {volume} {93}},\ \bibinfo {pages} {155113} (\bibinfo
  {year} {2016})}\BibitemShut {NoStop}%
\bibitem [{\citenamefont {Louvet}\ \emph {et~al.}(2016)\citenamefont {Louvet},
  \citenamefont {Carpentier},\ and\ \citenamefont {Fedorenko}}]{Louvet-2016}%
  \BibitemOpen
  \bibfield  {author} {\bibinfo {author} {\bibfnamefont {T.}~\bibnamefont
  {Louvet}}, \bibinfo {author} {\bibfnamefont {D.}~\bibnamefont {Carpentier}},
  \ and\ \bibinfo {author} {\bibfnamefont {A.~A.}\ \bibnamefont {Fedorenko}},\
  }\href@noop {} {\bibfield  {journal} {\bibinfo  {journal} {arXiv preprint
  arXiv:1605.02009}\ } (\bibinfo {year} {2016})}\BibitemShut {NoStop}%
\bibitem [{\citenamefont {Pixley}\ \emph
  {et~al.}(2016{\natexlab{b}})\citenamefont {Pixley}, \citenamefont {Huse},\
  and\ \citenamefont {Das~Sarma}}]{Pixley-2016}%
  \BibitemOpen
  \bibfield  {author} {\bibinfo {author} {\bibfnamefont {J.~H.}\ \bibnamefont
  {Pixley}}, \bibinfo {author} {\bibfnamefont {D.~A.}\ \bibnamefont {Huse}}, \
  and\ \bibinfo {author} {\bibfnamefont {S.}~\bibnamefont {Das~Sarma}},\ }\href
  {\doibase 10.1103/PhysRevX.6.021042} {\bibfield  {journal} {\bibinfo
  {journal} {Phys. Rev. X}\ }\textbf {\bibinfo {volume} {6}},\ \bibinfo {pages}
  {021042} (\bibinfo {year} {2016}{\natexlab{b}})}\BibitemShut {NoStop}%
\bibitem [{\citenamefont {Pixley}\ \emph
  {et~al.}(2016{\natexlab{c}})\citenamefont {Pixley}, \citenamefont {Huse},\
  and\ \citenamefont {Das~Sarma}}]{Pixley2}%
  \BibitemOpen
  \bibfield  {author} {\bibinfo {author} {\bibfnamefont {J.~H.}\ \bibnamefont
  {Pixley}}, \bibinfo {author} {\bibfnamefont {D.~A.}\ \bibnamefont {Huse}}, \
  and\ \bibinfo {author} {\bibfnamefont {S.}~\bibnamefont {Das~Sarma}},\ }\href
  {\doibase 10.1103/PhysRevB.94.121107} {\bibfield  {journal} {\bibinfo
  {journal} {Phys. Rev. B}\ }\textbf {\bibinfo {volume} {94}},\ \bibinfo
  {pages} {121107} (\bibinfo {year} {2016}{\natexlab{c}})}\BibitemShut
  {NoStop}%
\bibitem [{\citenamefont {Sbierski}\ \emph {et~al.}(2017)\citenamefont
  {Sbierski}, \citenamefont {Madsen}, \citenamefont {Brouwer},\ and\
  \citenamefont {Karrasch}}]{Sbierski-2017}%
  \BibitemOpen
  \bibfield  {author} {\bibinfo {author} {\bibfnamefont {B.}~\bibnamefont
  {Sbierski}}, \bibinfo {author} {\bibfnamefont {K.~A.}\ \bibnamefont
  {Madsen}}, \bibinfo {author} {\bibfnamefont {P.~W.}\ \bibnamefont {Brouwer}},
  \ and\ \bibinfo {author} {\bibfnamefont {C.}~\bibnamefont {Karrasch}},\
  }\href {\doibase 10.1103/PhysRevB.96.064203} {\bibfield  {journal} {\bibinfo
  {journal} {Phys. Rev. B}\ }\textbf {\bibinfo {volume} {96}},\ \bibinfo
  {pages} {064203} (\bibinfo {year} {2017})}\BibitemShut {NoStop}%
\bibitem [{\citenamefont {Pixley}\ \emph {et~al.}(2017)\citenamefont {Pixley},
  \citenamefont {Chou}, \citenamefont {Goswami}, \citenamefont {Huse},
  \citenamefont {Nandkishore}, \citenamefont {Radzihovsky},\ and\ \citenamefont
  {Das~Sarma}}]{Pixley-2017}%
  \BibitemOpen
  \bibfield  {author} {\bibinfo {author} {\bibfnamefont {J.~H.}\ \bibnamefont
  {Pixley}}, \bibinfo {author} {\bibfnamefont {Y.-Z.}\ \bibnamefont {Chou}},
  \bibinfo {author} {\bibfnamefont {P.}~\bibnamefont {Goswami}}, \bibinfo
  {author} {\bibfnamefont {D.~A.}\ \bibnamefont {Huse}}, \bibinfo {author}
  {\bibfnamefont {R.}~\bibnamefont {Nandkishore}}, \bibinfo {author}
  {\bibfnamefont {L.}~\bibnamefont {Radzihovsky}}, \ and\ \bibinfo {author}
  {\bibfnamefont {S.}~\bibnamefont {Das~Sarma}},\ }\href {\doibase
  10.1103/PhysRevB.95.235101} {\bibfield  {journal} {\bibinfo  {journal} {Phys.
  Rev. B}\ }\textbf {\bibinfo {volume} {95}},\ \bibinfo {pages} {235101}
  (\bibinfo {year} {2017})}\BibitemShut {NoStop}%
\bibitem [{\citenamefont {Gurarie}(2017)}]{Guararie-2017}%
  \BibitemOpen
  \bibfield  {author} {\bibinfo {author} {\bibfnamefont {V.}~\bibnamefont
  {Gurarie}},\ }\href {\doibase 10.1103/PhysRevB.96.014205} {\bibfield
  {journal} {\bibinfo  {journal} {Phys. Rev. B}\ }\textbf {\bibinfo {volume}
  {96}},\ \bibinfo {pages} {014205} (\bibinfo {year} {2017})}\BibitemShut
  {NoStop}%
\bibitem [{\citenamefont {Wilson}\ \emph {et~al.}(2017)\citenamefont {Wilson},
  \citenamefont {Pixley}, \citenamefont {Goswami},\ and\ \citenamefont
  {Das~Sarma}}]{Wilson-2017}%
  \BibitemOpen
  \bibfield  {author} {\bibinfo {author} {\bibfnamefont {J.~H.}\ \bibnamefont
  {Wilson}}, \bibinfo {author} {\bibfnamefont {J.~H.}\ \bibnamefont {Pixley}},
  \bibinfo {author} {\bibfnamefont {P.}~\bibnamefont {Goswami}}, \ and\
  \bibinfo {author} {\bibfnamefont {S.}~\bibnamefont {Das~Sarma}},\ }\href
  {\doibase 10.1103/PhysRevB.95.155122} {\bibfield  {journal} {\bibinfo
  {journal} {Phys. Rev. B}\ }\textbf {\bibinfo {volume} {95}},\ \bibinfo
  {pages} {155122} (\bibinfo {year} {2017})}\BibitemShut {NoStop}%
\bibitem [{\citenamefont {Wilson}\ \emph {et~al.}(2018)\citenamefont {Wilson},
  \citenamefont {Pixley}, \citenamefont {Huse}, \citenamefont {Refael},\ and\
  \citenamefont {Sarma}}]{Wilson-2018}%
  \BibitemOpen
  \bibfield  {author} {\bibinfo {author} {\bibfnamefont {J.~H.}\ \bibnamefont
  {Wilson}}, \bibinfo {author} {\bibfnamefont {J.~H.}\ \bibnamefont {Pixley}},
  \bibinfo {author} {\bibfnamefont {D.~A.}\ \bibnamefont {Huse}}, \bibinfo
  {author} {\bibfnamefont {G.}~\bibnamefont {Refael}}, \ and\ \bibinfo {author}
  {\bibfnamefont {S.~D.}\ \bibnamefont {Sarma}},\ }\href@noop {} {\bibfield
  {journal} {\bibinfo  {journal} {arXiv:1801.05438}\ } (\bibinfo {year}
  {2018})}\BibitemShut {NoStop}%
\bibitem [{\citenamefont {Syzranov}\ and\ \citenamefont
  {Radzihovsky}(2016)}]{Syzranov-2016}%
  \BibitemOpen
  \bibfield  {author} {\bibinfo {author} {\bibfnamefont {S.}~\bibnamefont
  {Syzranov}}\ and\ \bibinfo {author} {\bibfnamefont {L.}~\bibnamefont
  {Radzihovsky}},\ }\href@noop {} {\bibfield  {journal} {\bibinfo  {journal}
  {arXiv preprint arXiv:1609.05694}\ } (\bibinfo {year} {2016})}\BibitemShut
  {NoStop}%
\bibitem [{\citenamefont {De~Roeck}\ and\ \citenamefont
  {Huveneers}(2017)}]{DeRoeck-2017}%
  \BibitemOpen
  \bibfield  {author} {\bibinfo {author} {\bibfnamefont {W.}~\bibnamefont
  {De~Roeck}}\ and\ \bibinfo {author} {\bibfnamefont {F.~m.~c.}\ \bibnamefont
  {Huveneers}},\ }\href {\doibase 10.1103/PhysRevB.95.155129} {\bibfield
  {journal} {\bibinfo  {journal} {Phys. Rev. B}\ }\textbf {\bibinfo {volume}
  {95}},\ \bibinfo {pages} {155129} (\bibinfo {year} {2017})}\BibitemShut
  {NoStop}%
\bibitem [{\citenamefont {Wang}\ and\ \citenamefont {Chen}(2017)}]{wc2017}%
  \BibitemOpen
  \bibfield  {author} {\bibinfo {author} {\bibfnamefont {Y.}~\bibnamefont
  {Wang}}\ and\ \bibinfo {author} {\bibfnamefont {S.}~\bibnamefont {Chen}},\
  }\href {\doibase 10.1103/PhysRevA.95.053634} {\bibfield  {journal} {\bibinfo
  {journal} {Phys. Rev. A}\ }\textbf {\bibinfo {volume} {95}},\ \bibinfo
  {pages} {053634} (\bibinfo {year} {2017})}\BibitemShut {NoStop}%
\bibitem [{\citenamefont {Khemani}\ \emph {et~al.}(2017)\citenamefont
  {Khemani}, \citenamefont {Sheng},\ and\ \citenamefont {Huse}}]{Khemani-2017}%
  \BibitemOpen
  \bibfield  {author} {\bibinfo {author} {\bibfnamefont {V.}~\bibnamefont
  {Khemani}}, \bibinfo {author} {\bibfnamefont {D.~N.}\ \bibnamefont {Sheng}},
  \ and\ \bibinfo {author} {\bibfnamefont {D.~A.}\ \bibnamefont {Huse}},\
  }\href {\doibase 10.1103/PhysRevLett.119.075702} {\bibfield  {journal}
  {\bibinfo  {journal} {Phys. Rev. Lett.}\ }\textbf {\bibinfo {volume} {119}},\
  \bibinfo {pages} {075702} (\bibinfo {year} {2017})}\BibitemShut {NoStop}%
\bibitem [{\citenamefont {Wei\ss{}e}\ \emph {et~al.}(2006)\citenamefont
  {Wei\ss{}e}, \citenamefont {Wellein}, \citenamefont {Alvermann},\ and\
  \citenamefont {Fehske}}]{Weisse-2006}%
  \BibitemOpen
  \bibfield  {author} {\bibinfo {author} {\bibfnamefont {A.}~\bibnamefont
  {Wei\ss{}e}}, \bibinfo {author} {\bibfnamefont {G.}~\bibnamefont {Wellein}},
  \bibinfo {author} {\bibfnamefont {A.}~\bibnamefont {Alvermann}}, \ and\
  \bibinfo {author} {\bibfnamefont {H.}~\bibnamefont {Fehske}},\ }\href@noop {}
  {\bibfield  {journal} {\bibinfo  {journal} {Rev. Mod. Phys.}\ }\textbf
  {\bibinfo {volume} {78}},\ \bibinfo {pages} {275} (\bibinfo {year}
  {2006})}\BibitemShut {NoStop}%
\bibitem [{OSM()}]{OSM}%
  \BibitemOpen
  \href@noop {} {}\bibinfo {note} {See Online Supplemental Material for
  details.}\BibitemShut {Stop}%
\bibitem [{\citenamefont {Fehske}\ and\ \citenamefont
  {Schneider}(2008)}]{Fehske-2008}%
  \BibitemOpen
  \bibfield  {author} {\bibinfo {author} {\bibfnamefont {H.}~\bibnamefont
  {Fehske}}\ and\ \bibinfo {author} {\bibfnamefont {R.}~\bibnamefont
  {Schneider}},\ }\href@noop {} {\enquote {\bibinfo {title} {Aw (eds.).
  computational many-particle physics},}\ } (\bibinfo {year}
  {2008})\BibitemShut {NoStop}%
\bibitem [{\citenamefont {Devakul}\ and\ \citenamefont
  {Huse}(2017)}]{Devakul-2017}%
  \BibitemOpen
  \bibfield  {author} {\bibinfo {author} {\bibfnamefont {T.}~\bibnamefont
  {Devakul}}\ and\ \bibinfo {author} {\bibfnamefont {D.~A.}\ \bibnamefont
  {Huse}},\ }\href {\doibase 10.1103/PhysRevB.96.214201} {\bibfield  {journal}
  {\bibinfo  {journal} {Phys. Rev. B}\ }\textbf {\bibinfo {volume} {96}},\
  \bibinfo {pages} {214201} (\bibinfo {year} {2017})}\BibitemShut {NoStop}%
\bibitem [{\citenamefont {Jiang}(2012)}]{jiang_tunable_2012}%
  \BibitemOpen
  \bibfield  {author} {\bibinfo {author} {\bibfnamefont {J.-H.}\ \bibnamefont
  {Jiang}},\ }\href {\doibase 10.1103/PhysRevA.85.033640} {\bibfield  {journal}
  {\bibinfo  {journal} {Phys. Rev. A}\ }\textbf {\bibinfo {volume} {85}},\
  \bibinfo {pages} {033640} (\bibinfo {year} {2012})}\BibitemShut {NoStop}%
\bibitem [{\citenamefont {Sun}\ \emph {et~al.}(2012)\citenamefont {Sun},
  \citenamefont {Liu}, \citenamefont {Hemmerich},\ and\ \citenamefont
  {Sarma}}]{sun_topological_2012}%
  \BibitemOpen
  \bibfield  {author} {\bibinfo {author} {\bibfnamefont {K.}~\bibnamefont
  {Sun}}, \bibinfo {author} {\bibfnamefont {W.~V.}\ \bibnamefont {Liu}},
  \bibinfo {author} {\bibfnamefont {A.}~\bibnamefont {Hemmerich}}, \ and\
  \bibinfo {author} {\bibfnamefont {S.~D.}\ \bibnamefont {Sarma}},\ }\href
  {\doibase 10.1038/nphys2134} {\bibfield  {journal} {\bibinfo  {journal} {Nat.
  Phys.}\ }\textbf {\bibinfo {volume} {8}},\ \bibinfo {pages} {67} (\bibinfo
  {year} {2012})}\BibitemShut {NoStop}%
\bibitem [{\citenamefont {Dubček}\ \emph {et~al.}(2015)\citenamefont
  {Dubček}, \citenamefont {Kennedy}, \citenamefont {Lu}, \citenamefont
  {Ketterle}, \citenamefont {Soljačić},\ and\ \citenamefont
  {Buljan}}]{dubcek_weyl_2015}%
  \BibitemOpen
  \bibfield  {author} {\bibinfo {author} {\bibfnamefont {T.}~\bibnamefont
  {Dubček}}, \bibinfo {author} {\bibfnamefont {C.~J.}\ \bibnamefont
  {Kennedy}}, \bibinfo {author} {\bibfnamefont {L.}~\bibnamefont {Lu}},
  \bibinfo {author} {\bibfnamefont {W.}~\bibnamefont {Ketterle}}, \bibinfo
  {author} {\bibfnamefont {M.}~\bibnamefont {Soljačić}}, \ and\ \bibinfo
  {author} {\bibfnamefont {H.}~\bibnamefont {Buljan}},\ }\href {\doibase
  10.1103/PhysRevLett.114.225301} {\bibfield  {journal} {\bibinfo  {journal}
  {Phys. Rev. Lett.}\ }\textbf {\bibinfo {volume} {114}},\ \bibinfo {pages}
  {225301} (\bibinfo {year} {2015})}\BibitemShut {NoStop}%
\bibitem [{\citenamefont {Li}\ and\ \citenamefont
  {Sarma}(2015)}]{li_exotic_2015}%
  \BibitemOpen
  \bibfield  {author} {\bibinfo {author} {\bibfnamefont {X.}~\bibnamefont
  {Li}}\ and\ \bibinfo {author} {\bibfnamefont {S.~D.}\ \bibnamefont {Sarma}},\
  }\href {\doibase 10.1038/ncomms8137} {\bibfield  {journal} {\bibinfo
  {journal} {Nat. Commun.}\ }\textbf {\bibinfo {volume} {6}},\ \bibinfo {pages}
  {7137} (\bibinfo {year} {2015})}\BibitemShut {NoStop}%
\end{thebibliography}%


\begin{thebibliography}{1}
\bibitem{Pixley--2015} J. H. Pixley, Pallab Goswami, and S. Das Sarma, Phys. Rev. Lett. 115, 076601 (2015).
\bibitem{Pixley--2016} J. H. Pixley, David A. Huse, and S. Das Sarma, Phys. Rev. X 6, 021042 (2016).
\end{thebibliography}

%\newpage

\clearpage

\onecolumngrid

\setcounter{figure}{0}
\makeatletter
\renewcommand{\thefigure}{S\@arabic\c@figure}
\setcounter{equation}{0} \makeatletter
\renewcommand \theequation{S\@arabic\c@equation}
\renewcommand \thetable{S\@arabic\c@table}
\section*{ SUPPLEMENTAL MATERIAL: \\ Weyl semimetal to metal phase transitions driven by quasiperiodic potentials }
\setcounter{section}{0}

The Supplemental Material is organized as follows. In Sec.~I we discuss the scaling of the density of states with KPM expansion order and system size, for both the transition at $W_c$ and the miniband transition at $W_m$. In Sec.~II we discuss details of the behavior of the momentum-space inverse participation ratio (IPR) near these transitions. In Sec.~III we discuss how we determine the inverted semimetal phase. In Sec.~IV we present data on transport. Finally, in Sec.~V we present data at much stronger disorder, identifying the value at which Anderson localization sets in at the Weyl node energy.

\section{I: Scaling of the density of states}\label{criticalscaling}

\begin{figure}[h!]
\centering
\begin{minipage}{.3\textwidth}
\includegraphics[width=0.9\linewidth,angle=-90]{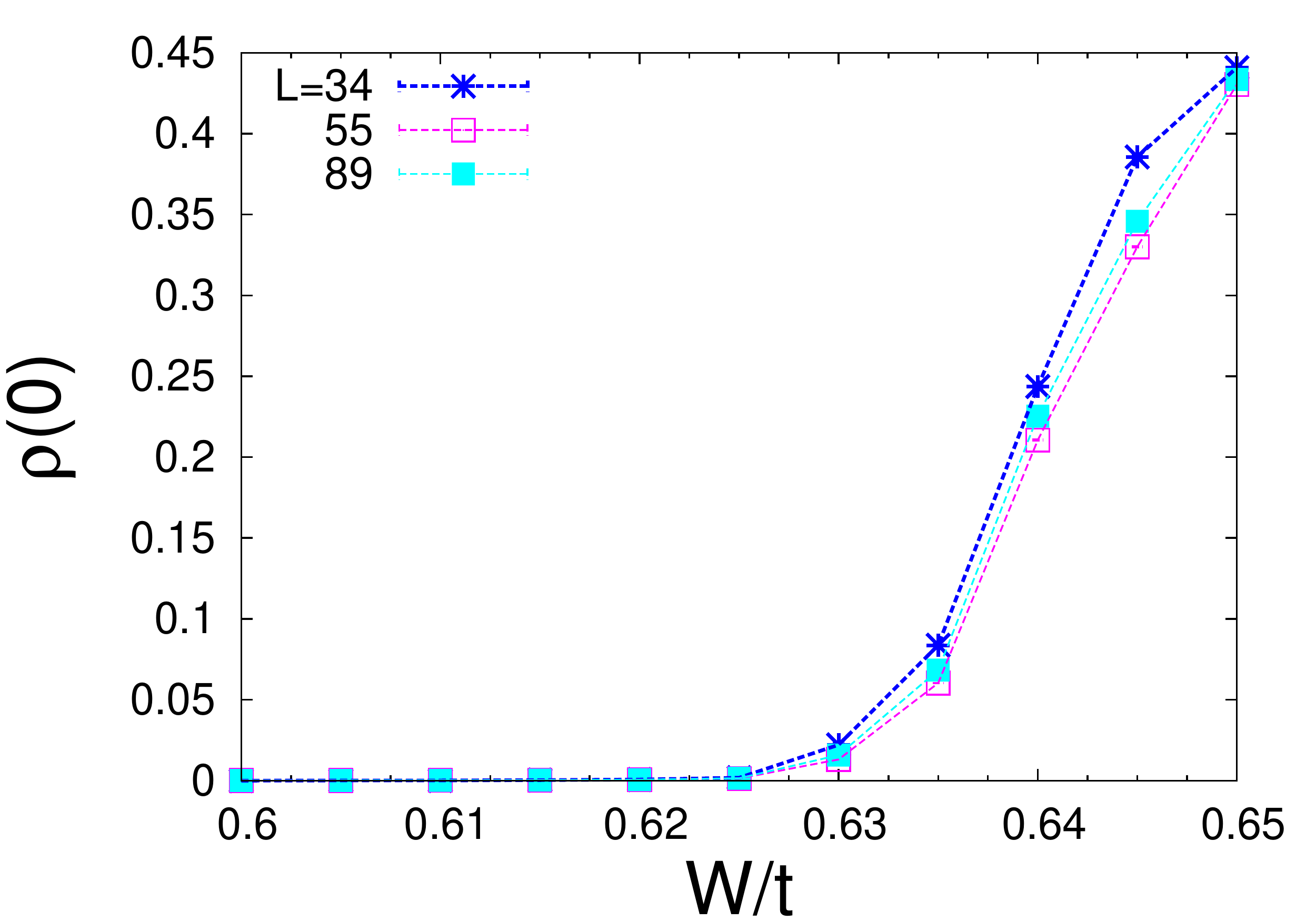}
\end{minipage}\hspace{5.pc}
\begin{minipage}{.3\textwidth}
\includegraphics[width=0.9\linewidth,angle=-90]{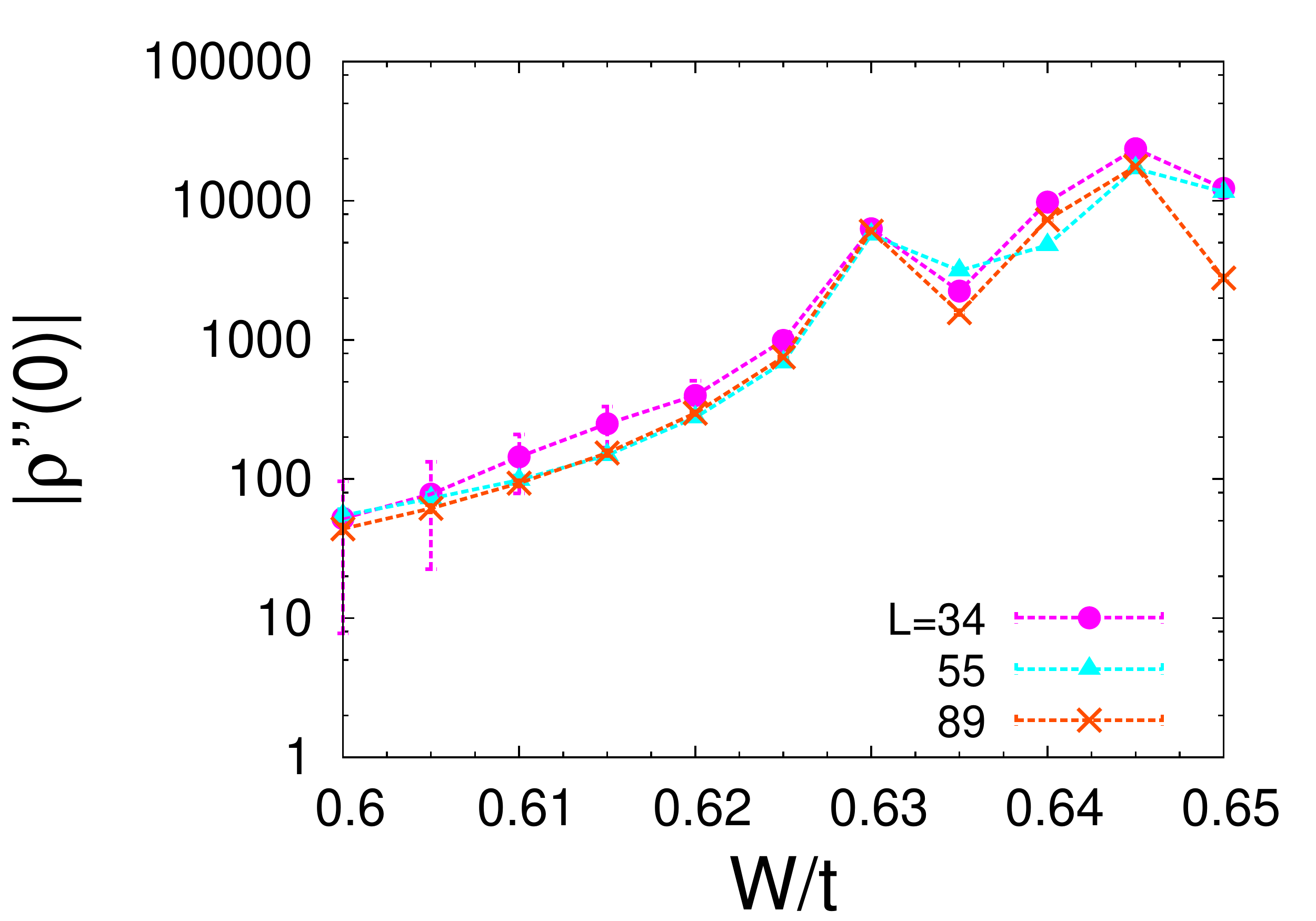}
\end{minipage}\hspace{5.pc}
\caption{(color online) System size dependence at fixed $N_C$ for $\rho(0)$ and $\rho''(0)$  for $N_C=2^{12}$. We find $\rho''(0)$ is essentially converged in $L$ in the semimetal phase and $\rho(0)$ is converged in $L$ in the diffusive metal phase for $N_C=2^{12}$.}
\label{fig:S1}
\end{figure}

\begin{figure}[h!]
\centering
\begin{minipage}{.3\textwidth}
\includegraphics[width=1.0\linewidth,angle=-90]{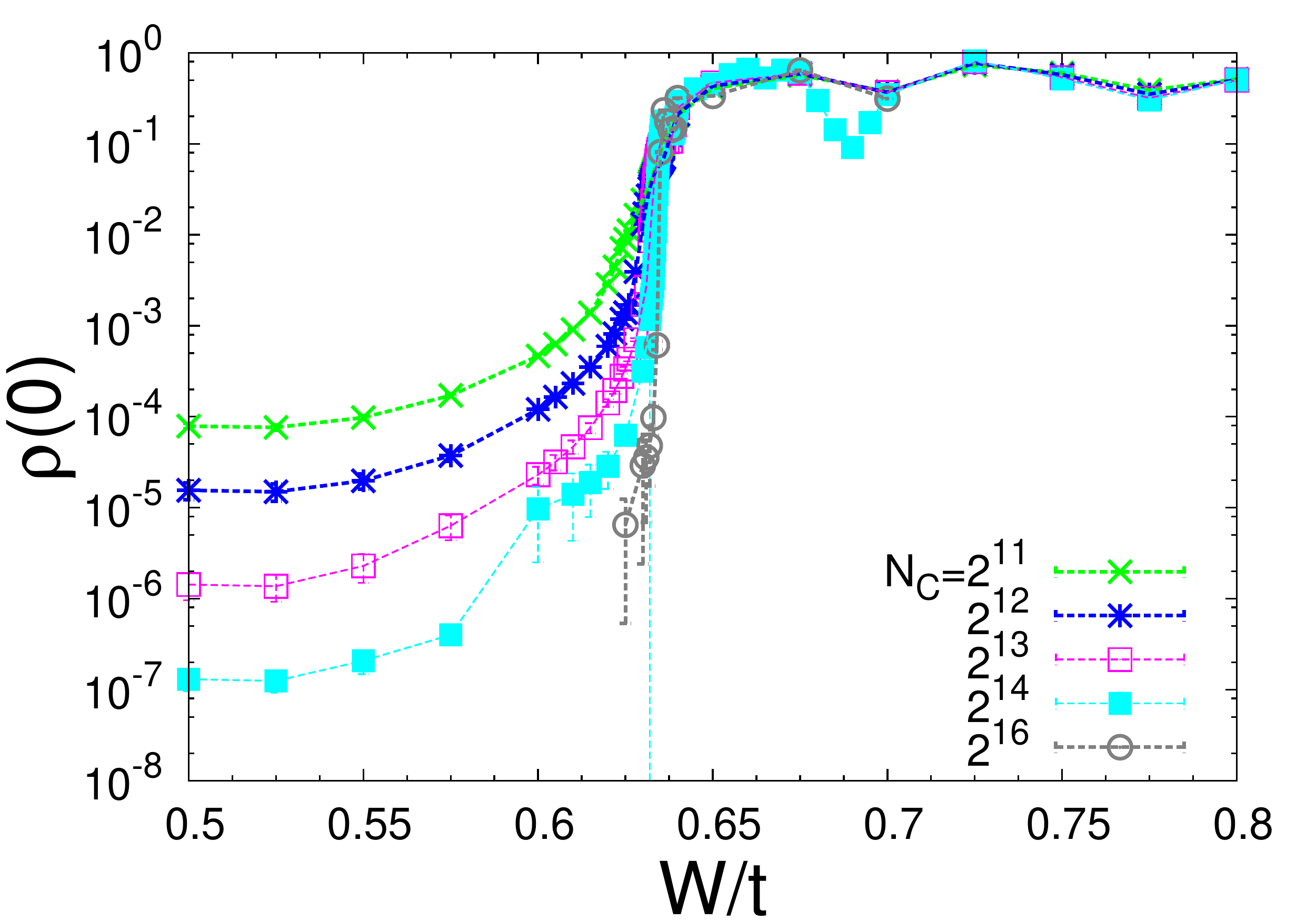}
\end{minipage}\hspace{5.pc}
\begin{minipage}{.3\textwidth}
\includegraphics[width=1.0\linewidth,angle=-90]{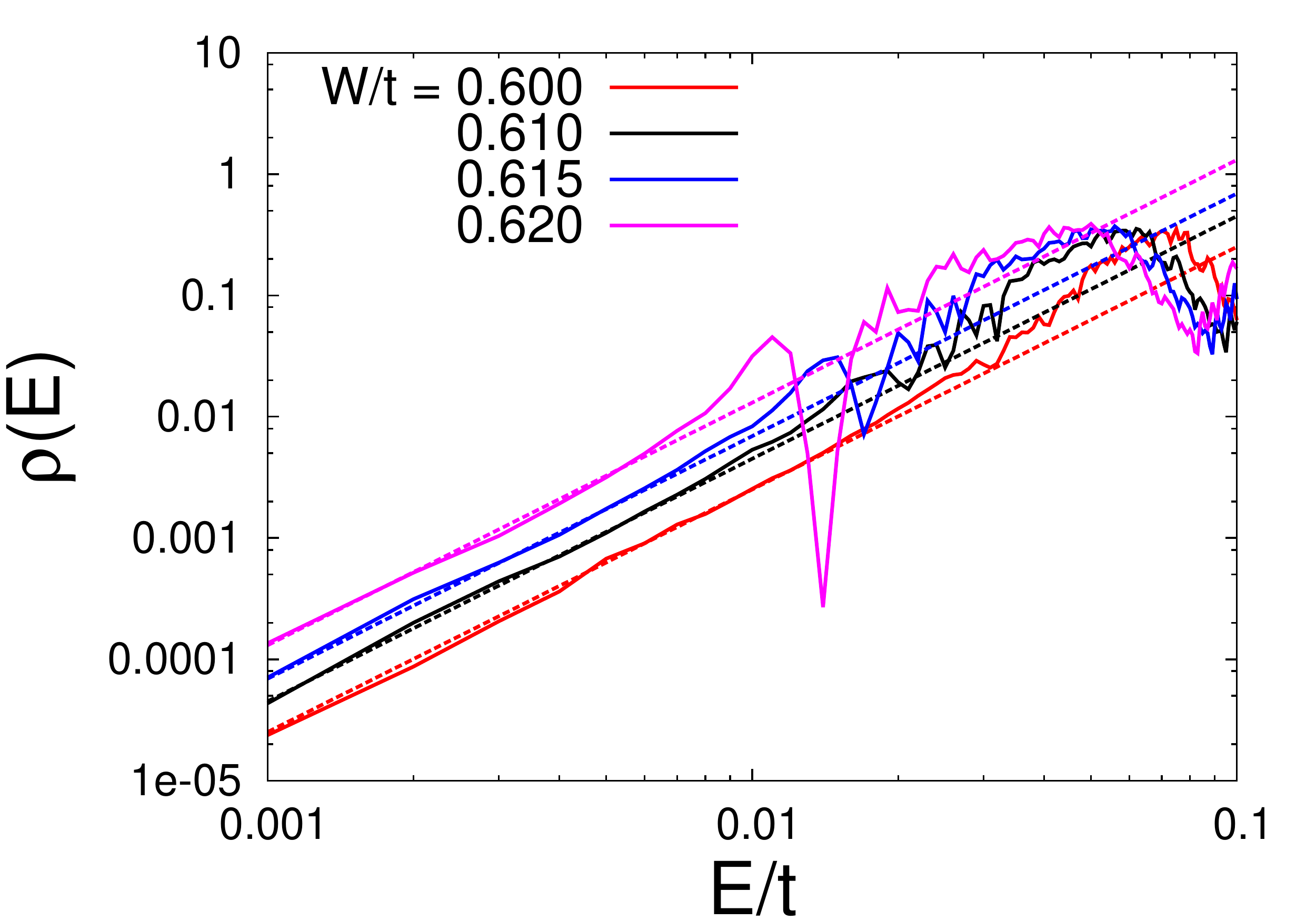}
\end{minipage}\hspace{2.pc}
\caption{(color online) (Left) Sharpness of the rise of the zero energy DOS near $W_c$. (Right) The energy dependence of $\rho(E)$ near the transition, the dashed lines are fits to $a E^2$ showing that the power law behavior does not change as the transition is approached, markedly distinct from the random problem. }
\label{fig:S2}
\end{figure}

In this section we show the finite-$L$ and finite-$N_C$ effects of the zero energy density of states (DOS) $\rho(0)$ and its second derivative $\rho''(0)$. Fig.~\ref{fig:S1} shows the scaling of these quantities with system size at fixed expansion order $N_C$; comparing to the $N_C$ dependence (main text, Fig. 2) shows that the data at moderately high $N_C$ are well-converged in $L$. Fig.~\ref{fig:S2} shows the $N_C$-dependence of $\rho(0)$ on a logarithmic scale; evidently the rise in the DOS becomes increasingly abrupt with expansion order, consistent with the possibility of a discontinuous DOS at the transition; another feature consistent with this possibility is the absence of any critical scaling window distinct from $\rho(E) \sim E^2$, shown in the right panel of that figure. 

\begin{figure}[h!]
\centering
\begin{minipage}{.225\textwidth}
\includegraphics[width=1.0\linewidth,angle=-90]{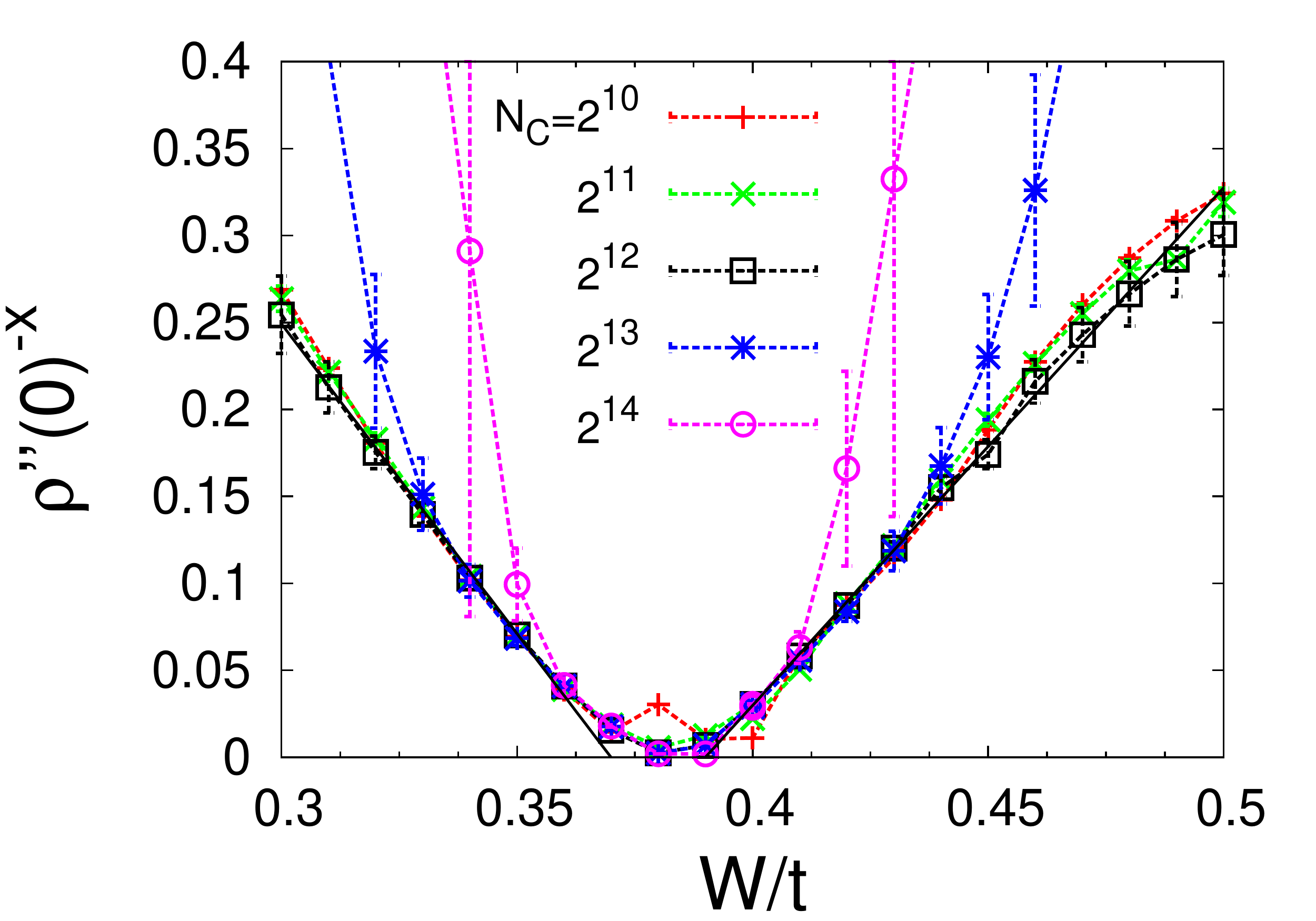}
\end{minipage}\hspace{4.pc}
\begin{minipage}{.225\textwidth}
\includegraphics[width=1.0\linewidth,angle=-90]{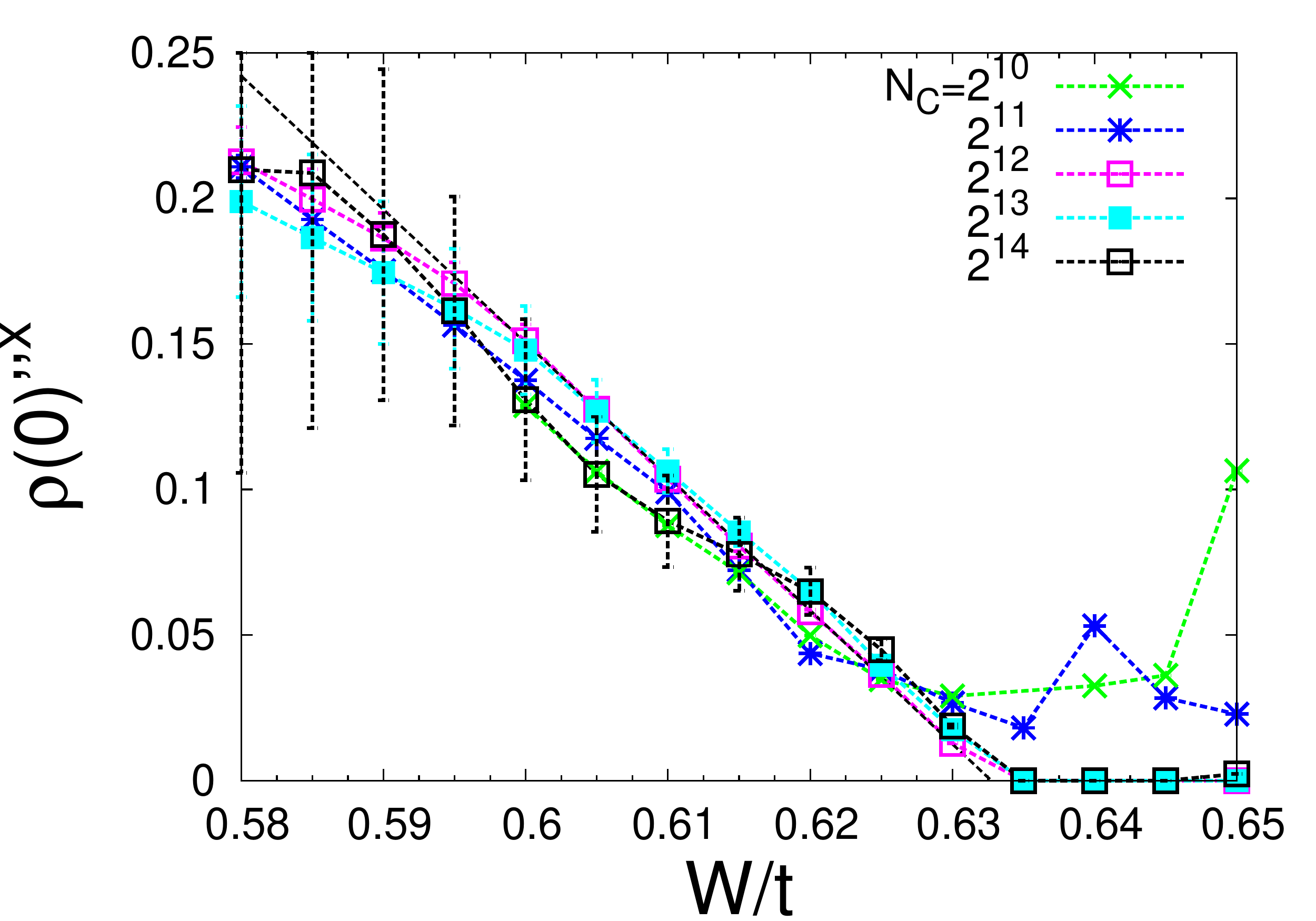}
\end{minipage}\hspace{4.pc}
\begin{minipage}{.225\textwidth}
\includegraphics[width=1.0\linewidth,angle=-90]{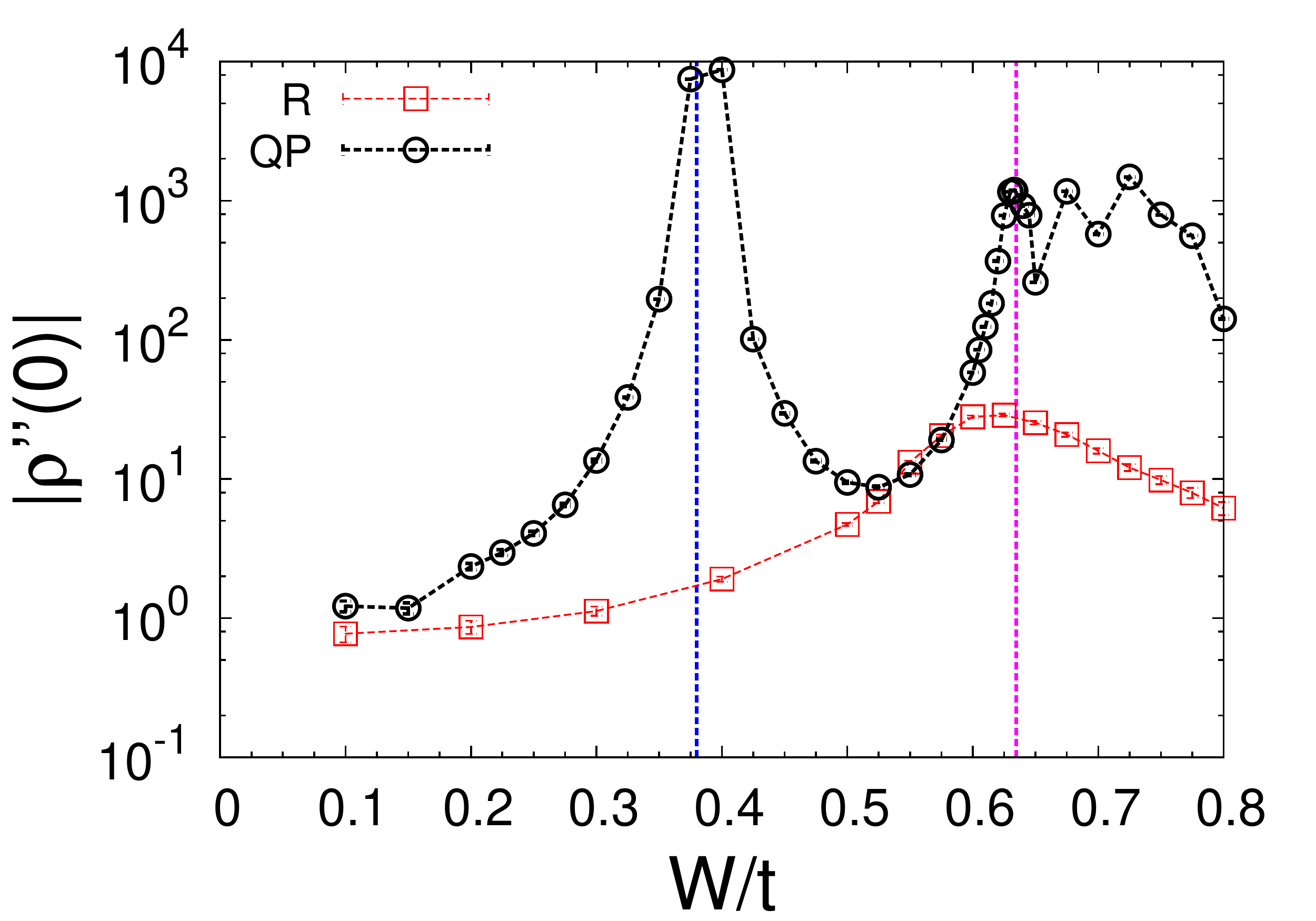}
\end{minipage}\hspace{4.pc}
\caption{(color online) Stability of the power law scaling regime near both transition in $\rho''(0)$ (Left) and (Center) for $L=89$. (Right) Comparison of the QP model and its randomized version (letting the phase $\phi$ be random at each site in the lattice). The dashed lines mark the miniband transition (blue) and the main transition (magenta).}
\label{fig:S3}
\end{figure}

\begin{figure}[h!]
\centering
\begin{minipage}{.3\textwidth}
\includegraphics[width=1.0\linewidth,angle=-90]{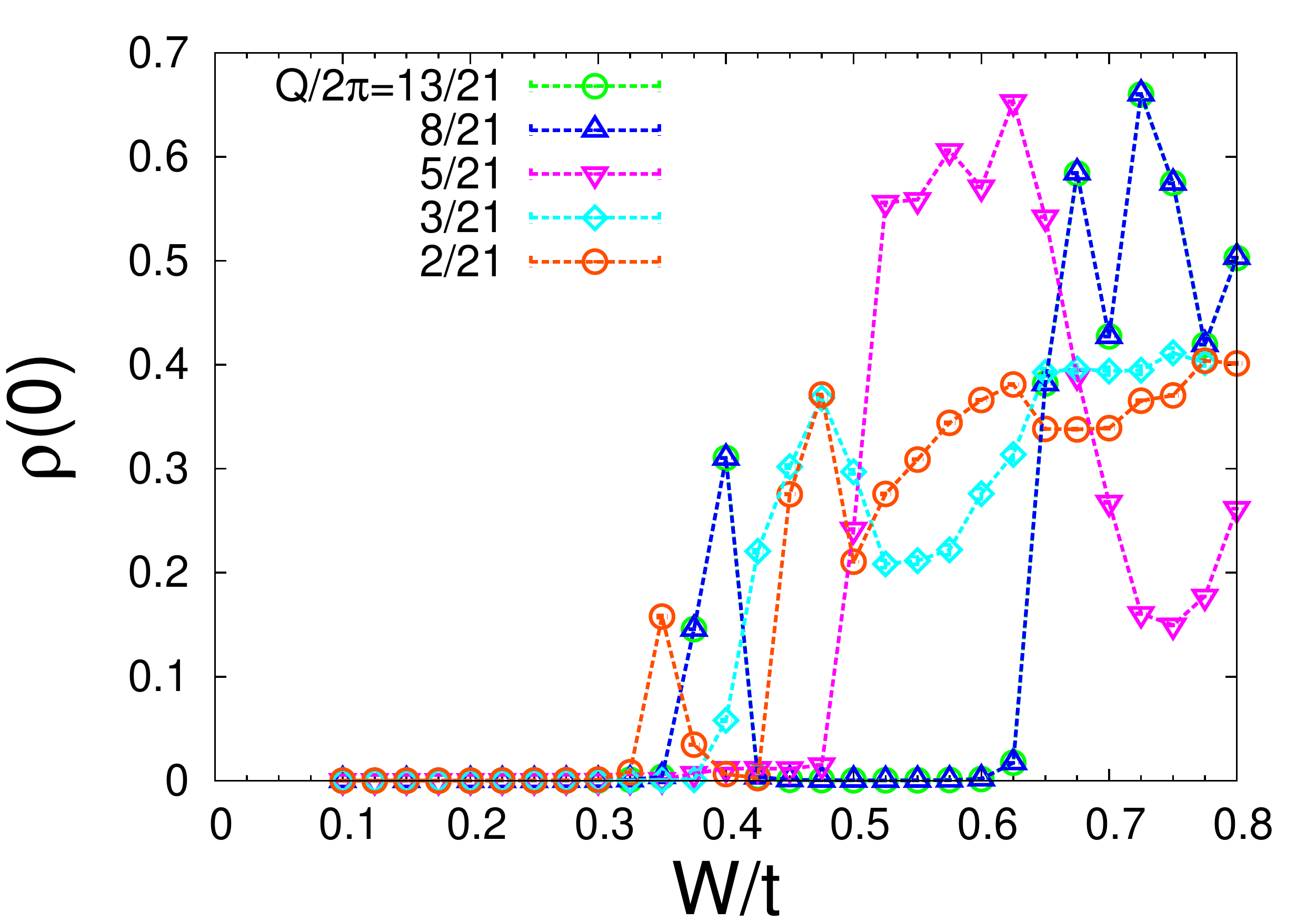}
\end{minipage}\hspace{6.pc}
\begin{minipage}{.3\textwidth}
\includegraphics[width=1.0\linewidth,angle=-90]{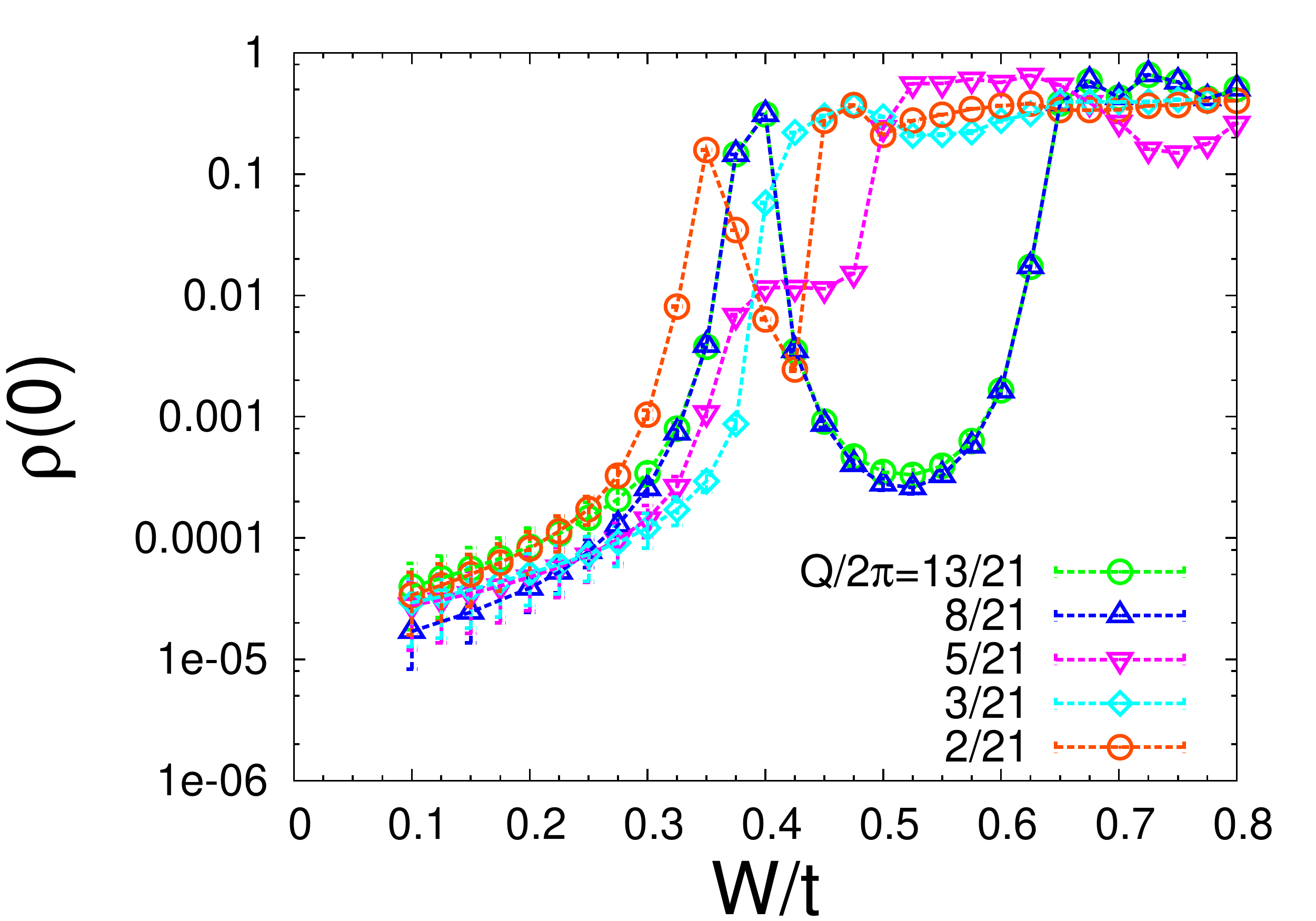}
\end{minipage}\hspace{4.pc}
\caption{(color online) Dependence of the location of the semimetal to metal transitions on our choice of the rational approximant wave vector $Q_L$ for $L=21$ and $N_C=2^{10}$. (Left) Linear scale of $\rho(0)$ versus $W$ and (Right) on a log linear scale. For $Q_L=2\pi F_{n-1}/F_n$ our results are equivalent to $Q_L=2\pi F_{n-2}/F_n$ due to the properties of Fibbonaci numbers this is just a shift of $2\pi$ to the potential. }
\label{fig:S3b}
\end{figure}

The scaling of $\rho''(0)$ at both miniband transitions closely follows that for the transition at $W_c$ (Fig.~\ref{fig:S3}), suggesting that all the semimetal-to-metal transitions we see are in the same universality class. The other features, such as the absence of a window with critical scaling of $\rho(E)$, are also common to all the cases we considered. Note also that the singularity of $\rho''(0)$ at the miniband transitions is even more pronounced than at $W_c$. We have also considered the dependence of our results on the choice of the wavevector $Q_L$, as shown in Fig.~\ref{fig:S3b}.

\section{II: Details of the momentum-space IPR}\label{iprk}

To assist in determining whether a phase is diffusive or ballistic, it is useful to look at the number of momentum (${\bf k}$) states that are ``participating'' in each eigenstate.
This is captured nicely by the inverse participation ratio (so-named because its inverse is in effect the number of participating states)
\begin{align}
  \mathcal I_k(E) = \left( \sum_{\mathbf k} |\psi_E(\mathbf k)|^2\right)^{-2} \sum_{\mathbf k} |\psi_E(\mathbf k)|^4,
\end{align}
where $\psi_E(\mathbf k)$ is energy eigenstate $E$ in the basis of plane wave solutions.
If this quantity is close to 1 and unchanging with $L$, the system size, then the eigenstates are localized in $\mathbf k$-space.
A decrease in this quantity with $L$ will represent non-ballistic (diffusive/localized) behavior, and in fact we see this in Fig.~1(d) of the main text for the model under consideration.

\begin{figure}
  \includegraphics[width=0.35\columnwidth]{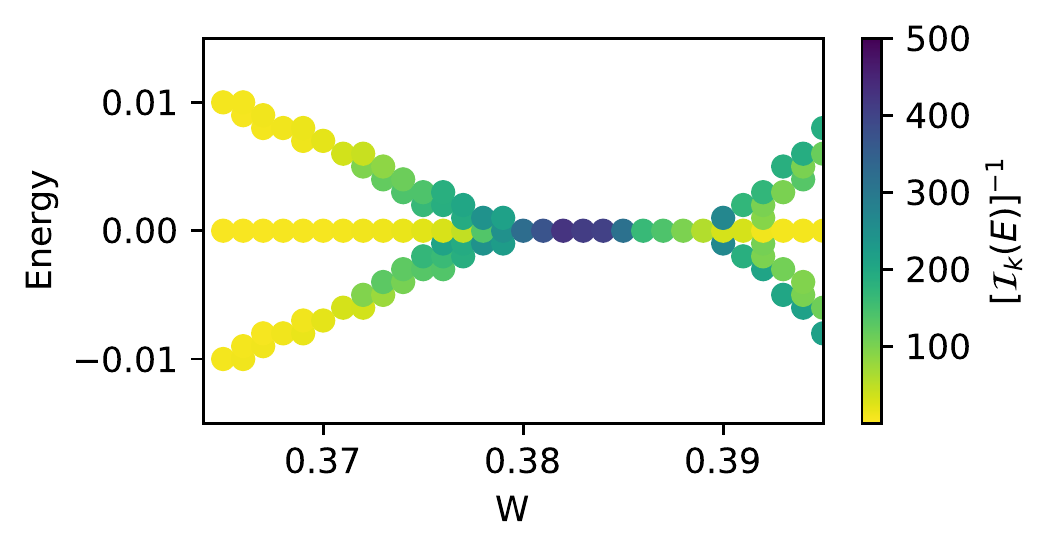}
  \includegraphics[width=0.3\columnwidth]{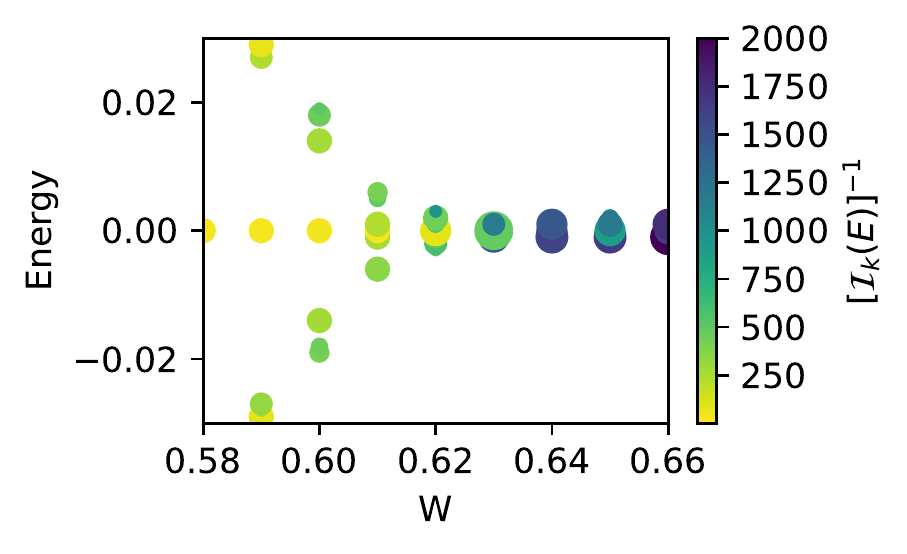}
  \includegraphics[width=0.25\columnwidth]{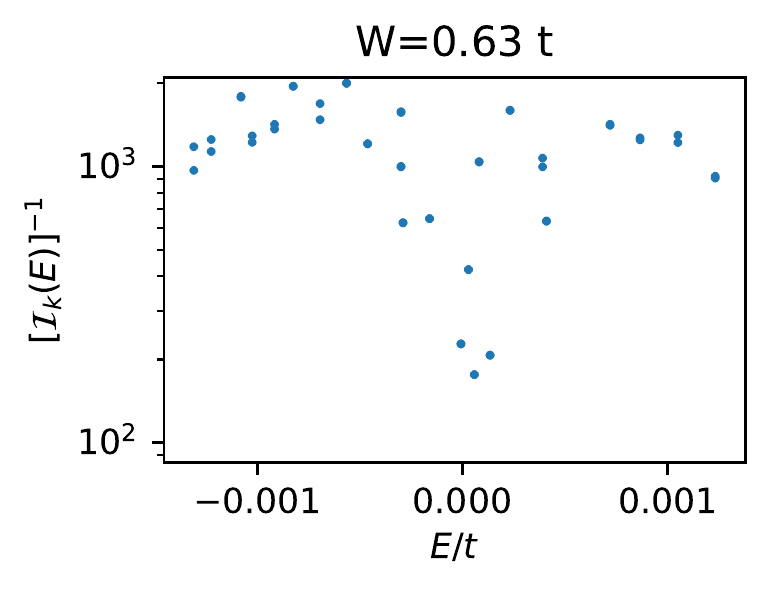}
  \caption{Plots of the $k$-space inverse participation ratio for $L=21$ for the eigenvectors with smallest eigenvalue in magnitude. (left) The mini-band transition from semimetallic to diffusive to inverted semimetal. (center) The transition from inverted semimetal to diffusive. (right) A slice of the $\mathcal I_k(E)$ data just after the transition from (inverted) semimetallic to diffusive for three realizations. We see considerable inhomogeneity in the data, all clustered closely around $E=0$.
  \label{fig:IPRk_L21}}
\end{figure}

For the numerical calculations involving $\mathcal I_k$, we are able to use the Lanczos algorithm to find low-lying energy state up to $L=21$. In Fig.~3(b) and Fig.~4(d) of the main text the result for the two transitions is given for $L=13$, here we show similar results for $L=21$ in Fig.~\ref{fig:IPRk_L21}.
Notably, the data for $\mathcal I_k(E)$ is rather inhomogenous after the final semimetallic to diffusive transition.

\section{III: Inverted semimetal}\label{sec:inverted-semimetal}

\begin{figure}
 \includegraphics[width=0.3\columnwidth]{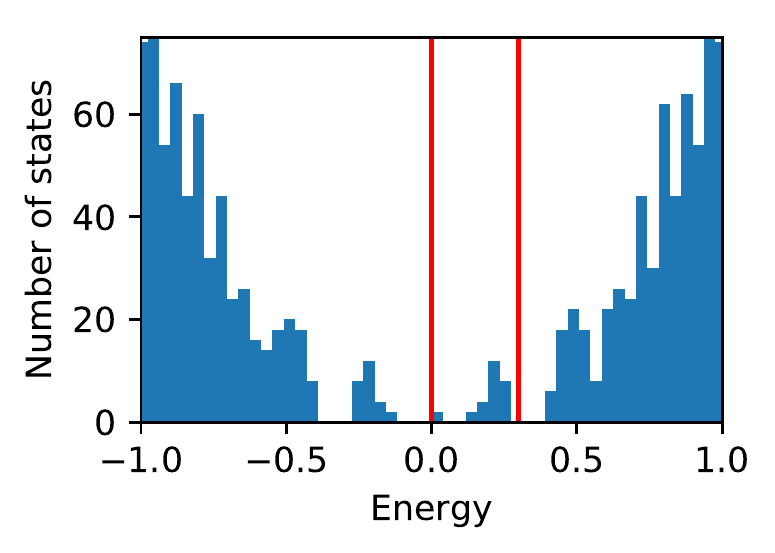}
 \includegraphics[width=0.5\columnwidth]{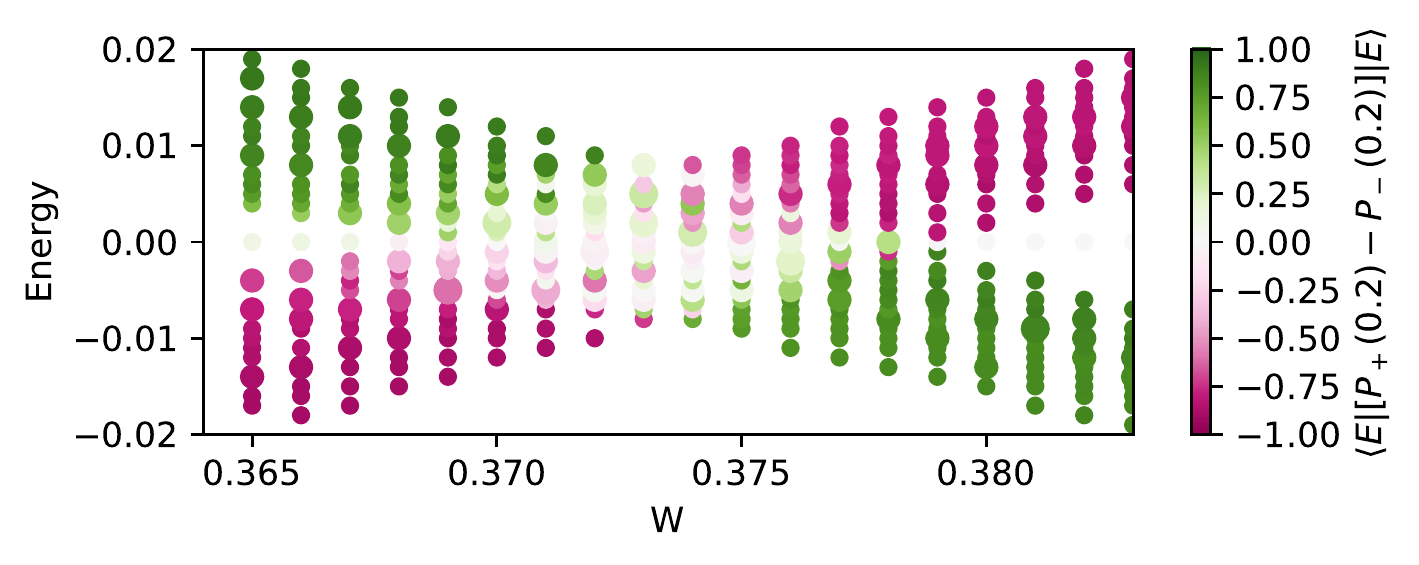}
 \caption{(left) At $W=0.2$ and $L=13$ we construct an operator $P_+(0.2)$ as the projection onto the eigenstates contained within the red lines. (right) At the miniband transition we see clearly positive and negative energies crossing and inverting the semimetal. The green states were initially positive energy Weyl states while the purple were initially negative energy Weyl states and the white rhombus where they cross represents the states hybridizing before separating out again. }
 \label{fig:inverstionOfBands}
\end{figure}

In the main text, we state that the semimetal inverts for certain values of the quasiperiodic potential strength $W$.
To show this, consider the positive energy miniband and construct its projection operator at a particular value of $W$ for a single realization: $P_+(W)$. Similarly, we can construct $P_-(W)$ for the negative energy miniband.
We first construct this operator using numerical data on $L=13$ at $W=0.2$: We take the states within the red lines of Fig.~\ref{fig:inverstionOfBands}(left) and construct $P_+(0.2)$ (similarly for $P_-(0.2)$).

Then as we scan in $W$, we perform exact diagonalization to obtain the eigenvectors $H(W)| E \rangle = E(W) | E \rangle $.
We then calculate the expectation values $\langle E | P_\pm(0.2) | E \rangle$ to see how much of these energy states live in each subspace.
The results are seen in Fig.~\ref{fig:inverstionOfBands}(right) where the color represents $\langle E | P_+(0.2) - P_-(0.2)| E \rangle$.
Even though this quantity is not the individual $P_\pm(0.2)$, the plot remains virtually unchanged if we instead plot $ P_\pm(0.2)$. 
This demonstrates that the semimetal inverts. Further, within the small diffusive range where they cross, the eigenstates mix so that no one state is fully within the positive or negative energy band (in line with the $\mathcal I_k$ data and the level spacing ratio $[r]$ data).

\section{IV: Transport}\label{transport}

We now discuss transport. As noted in the main text, neither of our methods for measuring transport gives conclusive results. The KPM method can be used to time-evolve an initially localized wavepacket to late times in large systems; however, it is insufficiently energy-resolved to pick out the leading behavior around the Weyl point. On the other hand, exact diagonalization permits one to energy-resolve but at the price of restricting our analysis to small system sizes $L = 13$, for which (by dimensional analysis) finite-size effects become important at timescales between $10$ and $100$ (in units of inverse hopping). Unfortunately, the corresponding energy scales correspond to the energy width of the near-transition states, so we lack a clear window between the physically relevant scales and those at which finite-size effects begin to dominate. 

Fig.~\ref{fig:S4} shows that for small disorder $W \approx 0.1$ the late-time transport is ballistic (because an appreciable fraction of the spectrum is ballistic) whereas for large disorder $W \approx 0.8$ the late-time transport is diffusive, as there are essentially no ballistic states. We are not able to see any sharp signature at $W = W_c$, this is because the vast majority of states  on both sides of the transition at $W_c$ are diffusive. 

\begin{figure}[h!]
\begin{minipage}{.25\textwidth}
\includegraphics[width=1.0\linewidth,angle=-90]{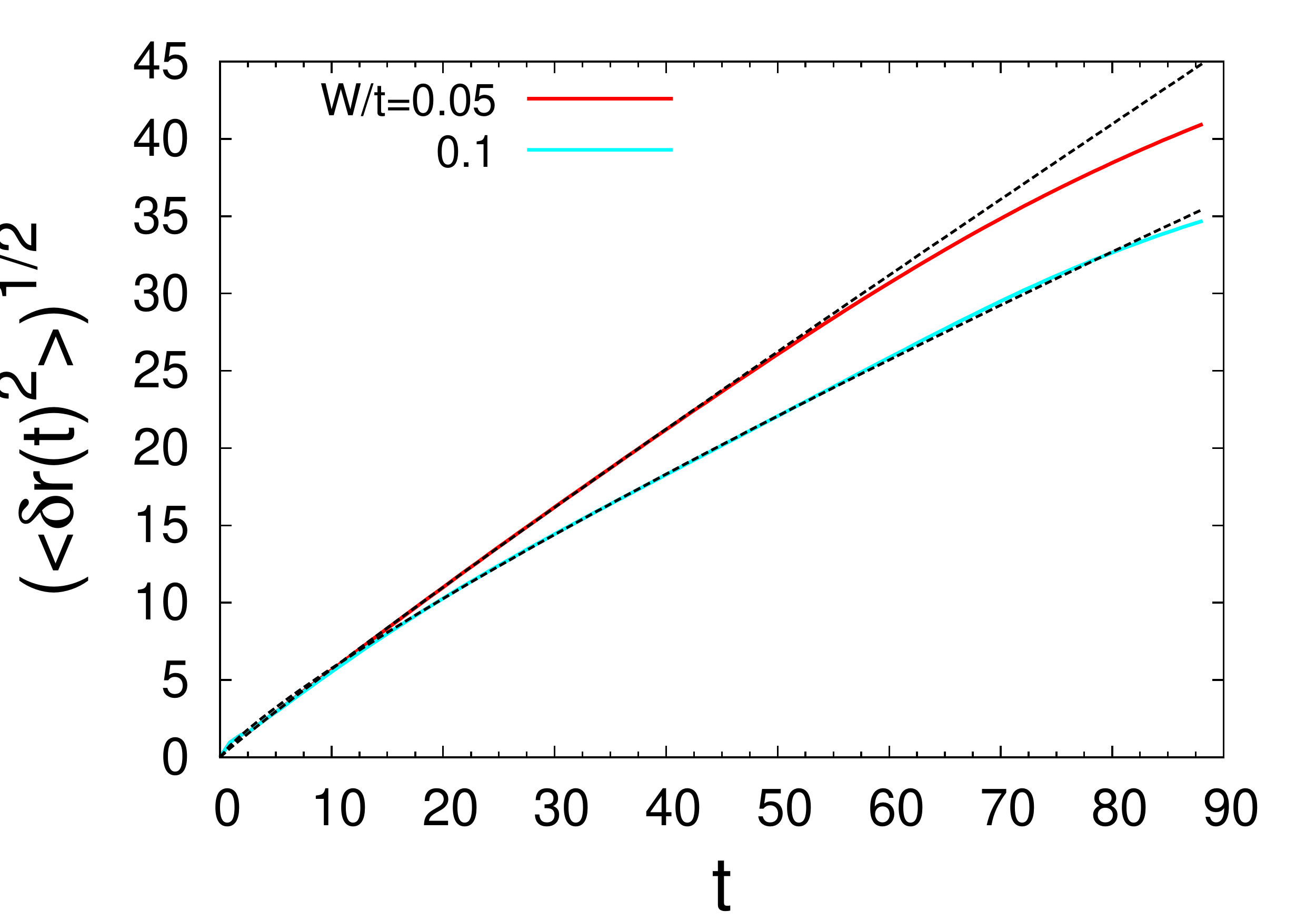}
\end{minipage}\hspace{6pc}
\begin{minipage}{.25\textwidth}
\includegraphics[width=1.0\linewidth,angle=-90]{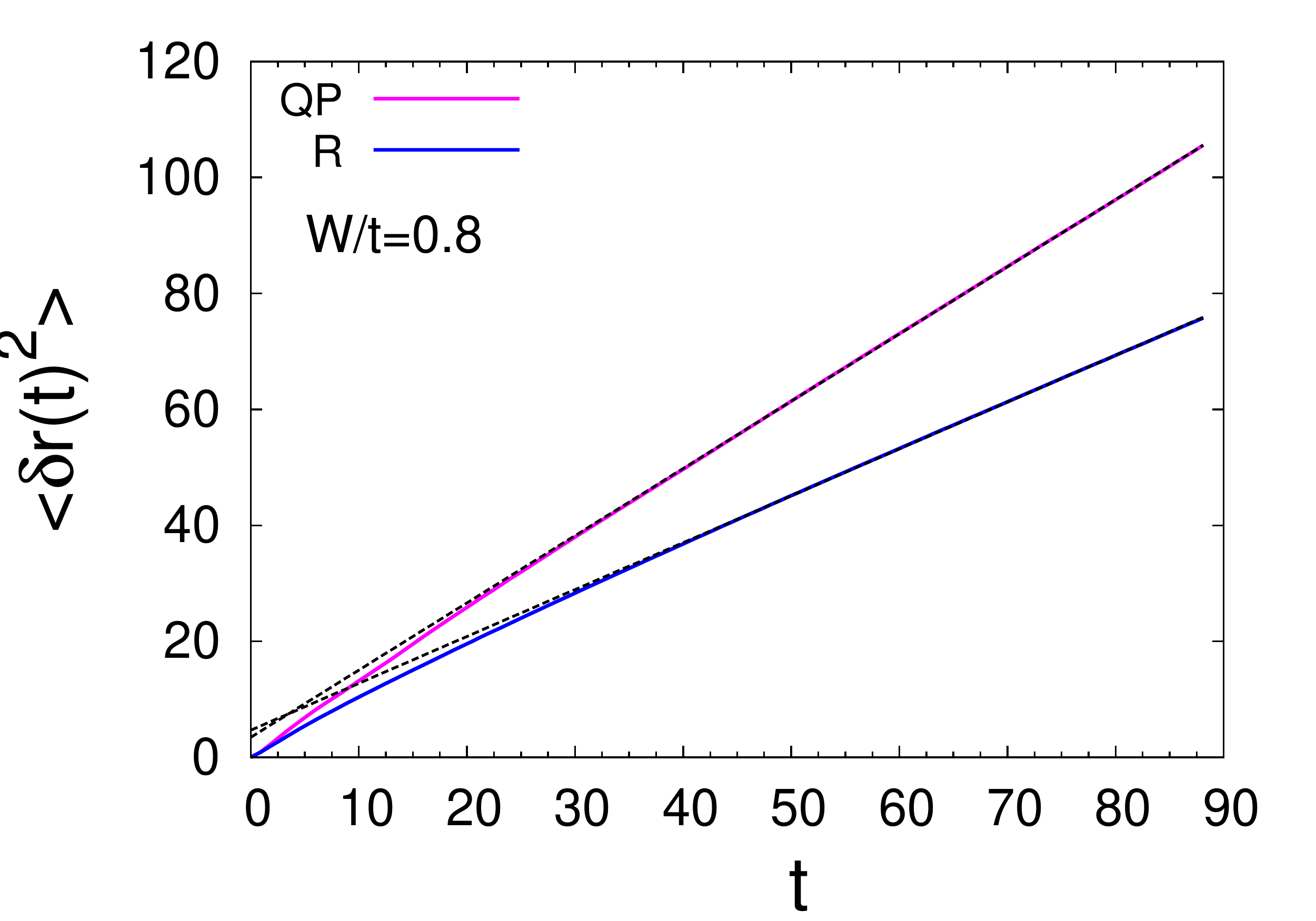}
\end{minipage}\hspace{3pc}
\caption{(color online) Wave packet dynamics starting with a wave function localized to a single site computed using the KPM on $L=89$. (Left) Ballistic scaling for weak quasiperiodic potential, the dashed lines are fits to $\langle \delta r(t)^2 \rangle \sim t^{\beta}$, which yields $\beta = 1.9$ and 1.7 for $W=0.05t$ and $0.1t$ respectively. (Right) Diffusive scaling in the diffusive metal phase comparing the QP and random (R) models, the dashed lines are fits to $\langle \delta r(t)^2 \rangle \sim D t$.}
\label{fig:S4}
\end{figure}

\begin{figure}[htbp]
\begin{center}
\begin{minipage}{.27\textwidth}
\includegraphics[width=1.0\linewidth]{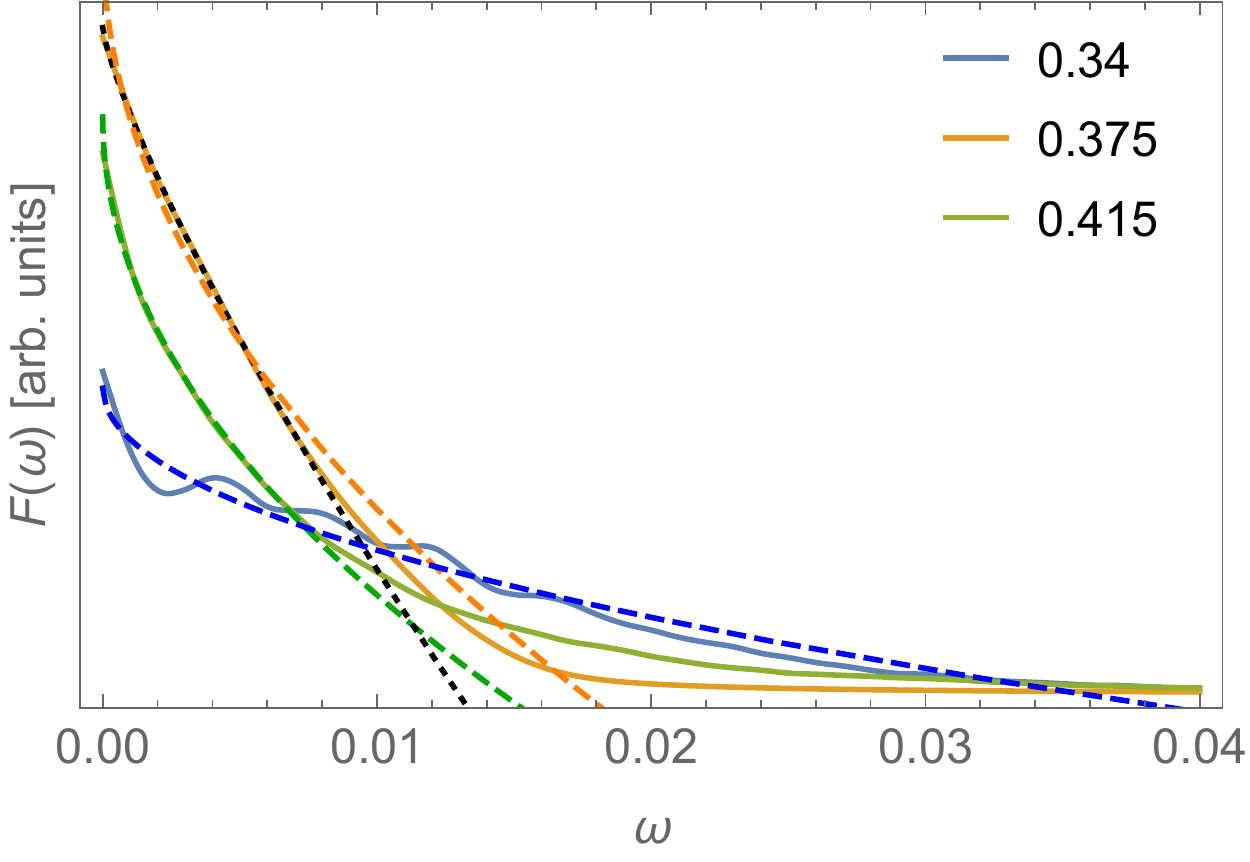}\hspace{3pc}
\end{minipage}
\begin{minipage}{.27\textwidth}
\includegraphics[width=1.0\linewidth]{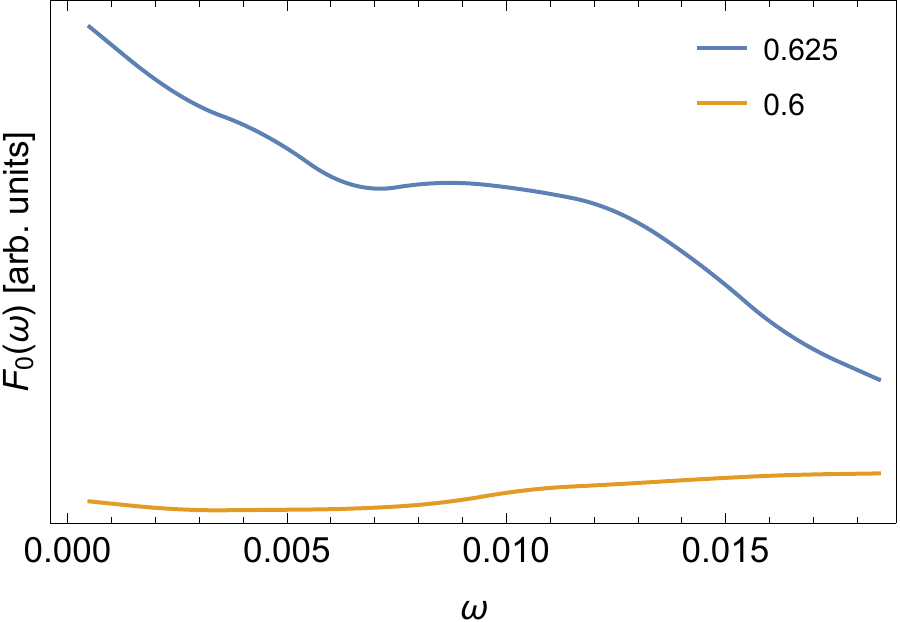}\hspace{3pc}
\end{minipage}
\begin{minipage}{.27\textwidth}
\includegraphics[width=1.0\linewidth]{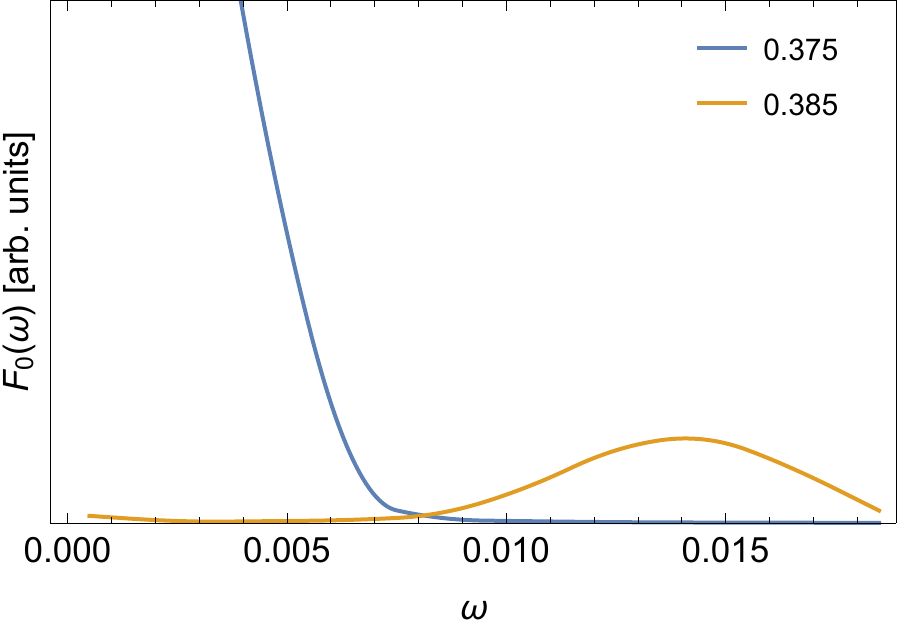}\hspace{3pc}
\end{minipage}
\caption{Frequency resolved spectral properties. Left: spectral functions at ``infinite temperature'' within the miniband, for various $W$. Dashed lines are fits to diffusive behavior, $F(\omega) = a + b \sqrt{\omega}$, which is consistent only for $W = 0.415$; the ballistic form, $F(\omega) = a + b \omega^2$, manifestly does not fit any of these curves. Dotted black line is a fit for $W = 0.375$ to the form $F(\omega) = a + b \omega^{0.8}$. Middle, right: zero-temperature spectral functions, for chemical potential at the Weyl point and various $W$.}
\label{spectral2}
\end{center}
\end{figure}

To study energy-resolved transport, we investigate the local dynamic structure factor, which is the Fourier transform of the density autocorrelation function, $\langle n_i(t) n_i(0) \rangle$. The structure factor, denoted $F(\omega)$, is

\begin{equation}
F(\omega) \propto \frac{1 - e^{-\beta \omega}}{\omega} \sum_{i, mn} p_m (1 - p_n) |\langle m | \hat{n}_i | n \rangle|^2 \delta[\omega - (E_n - E_m)] \sim \frac{1 - e^{-\beta \omega}}{\omega} \sum_{i,mn} p_m (1 - p_n) |\psi_m(i)|^2 |\psi_n(i)|^2 \delta[\omega - (E_n - E_m)].
\end{equation}
where $m, n$ are eigenstates and $p_m, p_n$ their occupation numbers. When the miniband is well-formed (for instance, near the miniband transition), one can simplify this further by ignoring Pauli blocking, and setting $p_m$ to be some small constant for states in the miniband and zero for states outside it. This corresponds to exploring the behavior of a wavepacket projected onto the miniband (as discussed in the main text). Diffusion would imply that the autocorrelation function goes as $t^{-3/2}$ at long times, so $F(\omega) \sim \mathrm{constant} + \omega^{1/2}$, where $a, b$ are constants. Ballistic propagation, on similar dimensional grounds, would give $F(\omega) \sim \mathrm{constant} + \omega^2$, i.e., smooth behavior near zero frequency (up to logarithmic corrections). Numerical results for $W = 0.34$, $W = 0.375$, and $W = 0.415$  are shown in the left panel of Fig.~\ref{spectral2}. No clear signature is seen of the putative diffusive-to-ballistic transitions: rather, $F(\omega)$ is most consistent with diffusion at $W = 0.415$, whereas in the nominally ``diffusive'' phase at $W = 0.375$ it appears superdiffusive but sub-ballistic. For $W = 0.375$ we see $F(\omega) - F(0) \sim \omega^{0.8}$, which corresponds to $\langle n_i(t) n_i(0) \rangle \sim 1/t^{1.8}$. However, all these results are severely limited by finite-size effects for our system size $L = 13$.

Near the transition at $W_c$, the above approach does not work as there is no well-defined ``miniband'' that is well separated in energy from the rest of the band structure. Instead, we take the chemical potential to be at $E = 0$ and compute the $T = 0$ response, denoted $F_0(\omega)$; because of Pauli blocking this is dominated by states close to zero energy. $F_0(\omega)$ is plotted in Fig.~\ref{spectral2}; it is dominated by the behavior of the density of states near zero energy. It jumps at the semimetal-to-metal transition (thus constituting a signature of that transition within transport), but this jump can be inferred from the concomitant jump of $\rho(E)$ at the transition.

\section{V: Anderson localization at strong quasiperiodic potential}\label{anderson}

\begin{figure}[h!]
\centering
\begin{minipage}{.3\textwidth}
\includegraphics[width=1.0\linewidth,angle=-90]{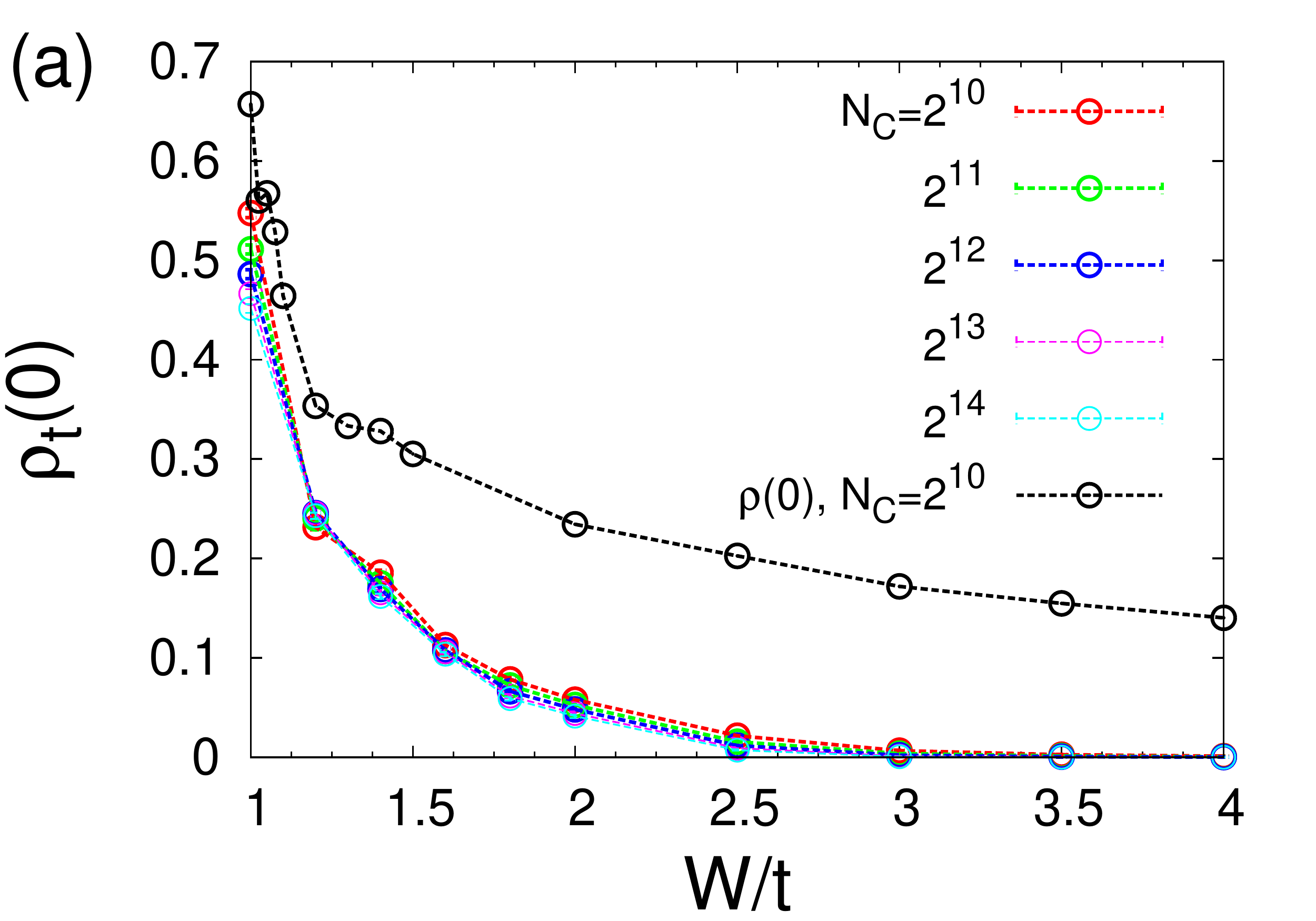}
\end{minipage}\hspace{5.5pc}
\centering
\begin{minipage}{.3\textwidth}
\includegraphics[width=1.0\linewidth,angle=-90]{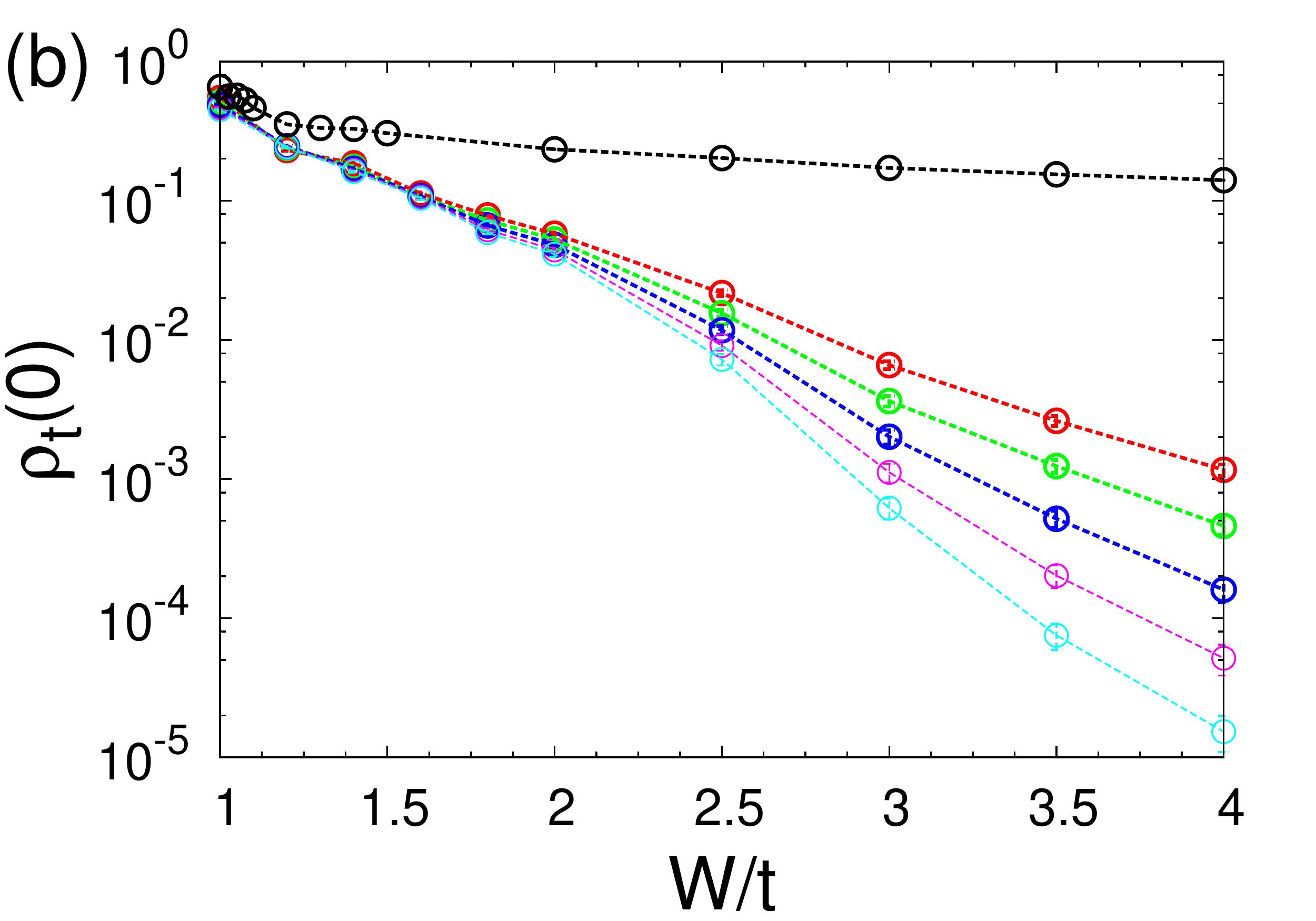}
\end{minipage}
\caption{(color online) Typical density states used to estimate the location of the Anderson localization transition, which we find occurs roughly at $W_l \approx 2.5t$ at $E=0$. (Left) Comparison of the average DOS to the typical DOS, displaying the typical DOS going to zero while the average remains finite. (Right) Log-linear scale showing the typical DOS develops a strong $N_C$ dependence upon entering the Anderson localized regime.}
\label{fig:S4}
\end{figure}

When the quasiperiodic potential is ramped up to values much higher than those addressed here, we expect Anderson localization to set in. A standard diagnostic of Anderson localization is the typical density of states, 
\begin{equation}
\rho_t(E) = \exp\left( \left [ \frac{1}{N_s}\sum_{i}^{N_s}\log \rho_i(E) \right] \right).
\end{equation}
Where we have introduced the local density of states $\rho_i(E) = \sum_{n,\alpha} |\langle n| i, \alpha \rangle|^2 \delta(E-E_n)$,  $N_s \ll V$ is a small number of sites that are randomly chosen and $[ \dots ]$ denotes a disorder average. We study the $N_C$ dependence of $\rho_t(0)$ to estimate the localization transition as done in Refs.~\cite{Pixley--2015,Pixley--2016}.
We find that the localization transition occurs at $W_l \approx 2.5t$ for $E=0$. 

%\bibliography{DSM_RR}

\end{document}